\documentclass[aps, prd, twocolumn, letterpaper, superscriptaddress, nofootinbib%
]{revtex4-1}

\usepackage{graphicx}
\usepackage{bbm,bm}
\usepackage{amssymb, amsmath,mathtools}
\usepackage[colorlinks=true, urlcolor=blue, linkcolor=black,%
  citecolor=blue]{hyperref}
\usepackage{enumerate}
\newcommand{\Nf}{N_\mathrm{f}}
\newcommand{\Eqref}[1]{Eq.~\eqref{#1}}
\DeclareMathOperator{\Tr}{Tr} 
\DeclareMathOperator{\STr}{STr} 
\DeclareMathOperator{\Span}{span}
\begin{document}

%===HEADER==============================================
\graphicspath{{./}{./plots/}}

\title{Fixed-point structure of low-dimensional relativistic fermion field theories: Universality classes and emergent symmetry}

\author{Friedrich Gehring}
\email{friedrich.gehring@uni-jena.de}
\affiliation{Theoretisch-Physikalisches Institut, Abbe Center of Photonics, 
Friedrich-Schiller-Universit{\"a}t Jena, Max-Wien-Platz 1, 07743 Jena, Germany}
\author{Holger Gies}
\email{holger.gies@uni-jena.de}
\affiliation{Theoretisch-Physikalisches Institut, Abbe Center of Photonics, 
Friedrich-Schiller-Universit{\"a}t Jena, Max-Wien-Platz 1, 07743 Jena, Germany}
\author{Lukas Janssen}
\email{lukasj@sfu.ca}
\affiliation{Department of Physics, Simon Fraser University, Burnaby, British Columbia V5A 1S6, Canada}
%=======================================================

%%%%%%%%%%%%%%%%%%%%%%%%%%%%%%%%%%%%%%%%%

\begin{abstract}
We investigate a class of relativistic fermion theories in $2<d<4$
space-time dimensions with continuous chiral $\mathrm U(\Nf) \times
\mathrm U(\Nf)$ symmetry. This includes a number of well-studied
models, e.g., of Gross-Neveu and Thirring type, in a unified
framework. Within the limit of pointlike interactions, the RG flow of
couplings reveals a network of interacting fixed points, each of which
defines a universality class. A subset of fixed points are ``critical
fixed points'' with one RG relevant direction being candidates for
critical points of second-order phase transitions. Identifying
invariant hyperplanes of the RG flow and classifying their
attractive/repulsive properties, we find evidence for emergent higher
chiral symmetries as a function of $\Nf$.  For the case of the
Thirring model, we discover a new critical flavor number that
separates the RG stable large-$\Nf$ regime from an intermediate-$\Nf$
regime in which symmetry-breaking perturbations become RG
relevant. This new critical flavor number has to be distinguished from
the chiral-critical flavor number, below which the Thirring model is
expected to allow spontaneous chiral symmetry breaking, and its
existence offers a resolution to the discrepancy between previous
results obtained in the continuum and the lattice Thirring models.
Moreover, we find indications for a new feature of universality:
details of the critical behavior can depend on additional ``spectator
symmetries'' that remain intact across the phase
transition. Implications for the physics of interacting fermions on
the honeycomb lattice, for which our theory space provides a simple
model, are given.
\end{abstract}

\maketitle

%%%%%%%%%%%%%%%%%%%%%%%%%%%%%%%%%%%%%%%%%
\section{Introduction}
The universal behavior of matter near the points of continuous phase
transitions constitutes one of the most intriguing phenomena in
statistical physics. Explaining universality was the great success of
renormalization group (RG) theory, which has by now become one of our
basic tools to understand systems with many interacting degrees of
freedom. Systems near criticality can fall into universality classes which
are characterized by only a few basic
properties, independent of the microscopic interactions between the
particles. In simple bosonic systems, the general characteristics that
determine the universality class are well established: the system's
dimensionality, the symmetry of the order parameter, and the presence
or absence of sufficiently long-ranged interactions.
The reason for this simplicity for a large class of purely bosonic 
theories lies in the existence of just one critical RG fixed point,
determining the critical behavior of all these theories within the
corresponding theory space~\cite{Herbut2007}.

In systems with fermions, however, the situation can be more complex,
and the question of the defining properties of the universality
classes could be more subtle. Some issues that arise in these systems
are the following: Is the above list exhaustive or can one find two
systems with the same dimensionality, symmetry of the order parameter,
and range of interaction, but \textit{different} critical behavior?
Does, for instance, the critical behavior depend on additional
(``spectator'') symmetries that do not take part in the symmetry
breaking pattern? Can there be more than one critical RG fixed point
in the same theory space? In fermionic systems, different types of
interactions often lead to various possible pairing mechanisms, which
allow excitations with new quantum numbers, and---upon condensation---lead to 
new collective modes. These new modes can be scalar fields
as in the BCS theory of superconductivity \cite{NambuJonaLasinio1961},
but other types such as vector modes are equally well possible
\cite{JanssenGies2012}. Furthermore, fermionic self-interactions are
inherently related to Fierz identities, with the help of which we can
always rewrite any four-fermion term as a linear combination of a
different set of four-fermion interactions. Focusing on only one
particular interaction channel and neglecting all others within a
single-channel approximation, as is usually done in mean-field
approaches, involves an ambiguity which may substantially affect the
validity of the approximation \cite{JaeckelWetterich2003,
  JanssenGies2012}. In a RG approach, instead, one should incorporate
\textit{all} interaction channels that are invariant under the
symmetry of a given system, and let the dynamics decide which one
becomes dominant. At the same time, this allows to study a whole class
of theories in a particular theory space and to investigate their
decomposition into universality classes.

In this paper, we present a class of relativistic fermion field
theories in $2<d<4$ space-time dimensions, which allows to study these
and related questions. In particular, we will investigate the space of
theories with Lorentz, continuous chiral $\mathrm U(\Nf) \times
\mathrm U(\Nf)$, and a set of discrete symmetries, with $\Nf$ being
the number of four-component fermion flavors.  This includes the
ubiquitous 3d Gross-Neveu models \cite{ RosensteinWarrPark1989,
  RosensteinYuKovner1993, Gracey1991,
  VasilievDerkachovKivelStepanenko1993, HandsKocicKogut1993,
  RosaVitaleWetterich2001, GiesJanssenRechenbergerScherer2010,
  BraunGiesScherer2011, ChandrasekharanLi2012}, as well as the 3d
Thirring model \cite{Parisi1975, DelDebbioHandsMehegan1997,
  ChristofiHandsStrouthos2007, Chandrasekharan2010,
  ChandrasekharanLi2012-b, GiesJanssen2010, JanssenGies2012,
  Mesterhazy2012}, both of which have been used as testing grounds to
study nonperturbative phenomena in strongly-coupled fermion field
theories, such as chiral symmetry breaking and nonperturbative
renormalizability. Lately,
these systems receive revived
attention as effective models describing the physics of condensed
matter systems that incorporate fermionic excitations with
relativistic dispersion relation, in particular graphene \cite{
  Herbut2006, HerbutJuricicRoy2009, JanssenHerbut2014,
  HandsStrouthos2008, CortijoGuineaVozmediano2012} and the surface
states of topological insulators \cite{VafekVishwanath2014}. Since
interactions in graphene are strong \cite{EliasEtAl2011}, the question
of possible quantum transitions from the semimetallic into different
Mott insulating \cite{Herbut2006, RaghuQiHonerkampZhang2008} or
superconducting \cite{RoyJuricicHerbut2013} phases has extensively
been investigated previously. We here take a somewhat different
viewpoint: Instead of focusing on the various possible infrared (IR)
phases \cite{RyuMudryHouChamon2009}, we target at the ultraviolet (UV)
structure of our effective-theory space. With the aid of functional RG
techniques, we map out the fixed points and the accompanying relevant
and irrelevant directions. A fixed point with exactly one relevant
direction corresponds to a possible second-order phase transition
whose critical behavior it governs. We demonstrate that in the present
theory space multiple such critical fixed points may exist, each one
of them defining its own universality class. Different theories with
identical field content and microscopic symmetries may therefore be in
the domain of attraction of \textit{different} critical fixed points
and thus be in different universality classes---even though
symmetry-breaking patterns and critical degrees of freedom may be
completely the same.

The theory space we propose also represents a simple example to study
the possibility of dynamical emergence of symmetry due to an
IR-attractive RG fixed point in a higher-symmetric subspace. Emergent
symmetry is a well-known phenomenon in various condensed-matter
systems, which often exhibit rotational symmetry at low energy, while
the microscopic Hamiltonian explicitly breaks (continuous) rotational
invariance. For lattice regularizations of rotational-invariant field
theories, dynamic enhancement of the rotation symmetry is essential in
order to avoid the need for fine tuning.  The possibility that Lorentz
symmetry, instead of at very high energies being explicitly realized,
could be emergent as a low-energy phenomenon is an old idea
\cite{ChadhaNielsen1983} and a key requirement for
Horava's approach to regularize quantum gravity \cite{Horava2009}. It
is also believed that the critical points in the graphene system are
Lorentz symmetric \cite{Herbut2006, HerbutJuricicRoy2009}.  Close to a
strong-coupling fixed point in (3+1)-dimensional conformal field
theory \cite{GohLutyNg2005} and in (2+1)-dimensional~\cite{Lee2007} 
and (3+1)-dimensional~\cite{jian2015}
fermionic lattice models supersymmetry could be
emergent. Emergent supersymmetry would be highly desirable for lattice
formulations of supersymmetric theories, which inevitably break at
least part of the supersymmetry on the microscopic level
\cite{Kaplan2004}. The emergence of enhanced internal symmetries
has also been observed 
near a fermionic multicritical point with $\mathbbm{Z}_2 \times 
\mathrm{O}(2)$ symmetry \cite{Roy2011} and
in bosonic $\mathrm{O}(N_1)+\mathrm{O}(N_2)$
models \cite{Calabrese2003}, leading to the
seemingly paradoxical possibility that Goldstone modes can arise in
models with initially only discrete symmetries.

The $\mathrm U(\Nf) \times \mathrm U(\Nf)$ theory space we consider in
this work includes the higher-symmetric $\mathrm U(2\Nf)$ subspace,
which is the theory space of the continuum Thirring model
\cite{GiesJanssen2010, JanssenGies2012}. Already for this simplistic
example, the question whether or not lower-symmetric perturbations out
of this subspace are relevant in the sense of the RG appears to be
nontrivial. It turns out that it can actually depend on the number of
flavors $\Nf$: Within our approximation we find that for $\Nf > 6$ the
$\mathrm U(2\Nf)$-symmetric subspace is IR attractive, in accordance
with the large-$\Nf$ analysis \cite{DelDebbioHandsMehegan1997,
  ChristofiHandsStrouthos2007}, while it becomes IR repulsive for
$2\leq \Nf<6$. Due to an additional Fierz identity the $\Nf=1$ case is
special, and we again find that perturbations out of the $\mathrm
U(2\Nf)$ subspace are irrelevant, as in the large-$\Nf$ case. These
findings shed new light on previous simulation results that employ
lattice formulations that generically break parts of the microscopic
symmetries of the continuum theories~\cite{DelDebbioHandsMehegan1997,
  ChristofiHandsStrouthos2007, Chandrasekharan2010,
  ChandrasekharanLi2012-b, ChandrasekharanLi2012}.

In order to determine the RG flow of our theory space we use
Wetterich's functional RG equation \cite{Wetterich1993}. For a first
analysis, we confine ourselves to the study of pointlike fermionic
interactions which is similar in spirit to the quantitatively
successful derivative expansion for bosonic theories.  As we shall
show, this approximation is equivalent to the usual Wilsonian one-loop
RG and as such, will become exact to first order in $d=2+\epsilon$ for
all $\Nf$, or in any dimension $2<d<4$ for large $\Nf$.  For the
physically interesting case of small $\Nf$ directly in $d=3$, our
simple approximation---as the great majority of all other analytical
approaches in the nonperturbative domain---may be not sufficiently
controlled.  However, the use of the functional RG in the present case
has important advantages: first, the method provides multiple
systematic ways to straightforwardly improve the present simple
approximation used here, e.g, by incorporating momentum-dependent
vertices or by partial- or dynamical-bosonization techniques---all of
which are employed and advanced in recent studies of similar systems
\cite{ MetznerSalmhoferHonerkampMedenSchoenhammer2012,
  GieringSalmhofer2012, LeeStrackSachdev2013, GiesWetterich2002,
  Pawlowski2007, JanssenGies2012}.  Second, our approach allows to
derive a general formula for the one-loop flow of relativistic fermion
models, which should be directly applicable to systems with an
arbitrary number of interaction channels. We believe that this will be
of relevance to future investigation of more complex systems with
critical fermion interactions, e.g., in order to derive an effective
theory for electrons in the single- \cite{HerbutJuricicRoy2009} or
bilayer \cite{Vafek2010} graphene in 2+1 dimensions, or in the quantum
critical systems with quadratic band touching in 3+1 dimensions
\cite{HerbutJanssen2014, SavaryMoonBalents2014}.

The rest of the article is organized as follows: In the following 
section we define the theory space that we consider and discuss its 
symmetries. We
briefly introduce our method in Sec.~\ref{sec:frg} which we will use in 
Sec.~\ref{sec:general-4-fermi} to derive a general formula for the one-loop 
beta function of relativistic fermion systems with pointlike four-fermion
interactions. We discuss the flow equations for our system in 
Sec.~\ref{sec:flow-equations}. Section~\ref{sec:general-one-loop} is devoted to
general properties of the one-loop flow in four-fermion models. In 
Secs.~\ref{sec:fixed-points-Nf-1} and \ref{sec:fixed-points-Nf-general} we
discuss the fixed-point structure for $\Nf = 1$ and $\Nf \geq 2$, 
respectively. We give an outlook on possible phase transitions and
critical behavior in Sec.~\ref{sec:prospects-IR}, and conclude in 
Sec.~\ref{sec:conclusions}.

%%%%%%%%%%%%%%%%%%%%%%%%%%%%%%%%%%%%%%%%%%%%%%%%%%%%%%%%%%%%%%%%%%%%%%%%%%%%%%%%
\section{Fermion models with \texorpdfstring{$\mathrm{U}(\Nf)\times \mathrm{U}(\Nf)$}{U(Nf)xU(Nf)} symmetry} \label{sec:symmetry}
We are interested in the theory space of the $\mathrm U(\Nf) \times
\mathrm U(\Nf)$-symmetric Gross-Neveu model in $2<d<4$ space-time
dimensions, which may be defined by the microscopic action
\cite{BraunGiesScherer2011}
\begin{align} \label{eq:action-GN}
S_\mathrm{GN} = \int d^d x \left[ \bar \psi^a i\gamma_\mu \partial_\mu \psi^a + \frac{\bar g}{2\Nf} \left(\bar\psi^a \psi^a\right)^2 \right],
\end{align}
with space-time index $\mu = 0,\dots,d-1$ and ``flavor'' index
$a=1,\dots,\Nf$. Summation over repeated indices is implicitly
understood. We use a four-dimensional representation of the Clifford
algebra $\{ \gamma_\mu, \gamma_\nu\} = 2 \delta_{\mu\nu} \mathbbm
1_4$.  The Dirac conjugate is given by $\bar\psi = -i \psi^\dagger
\gamma_0$.  A possible application of this model is the system of
$\Nf$ species of fermions on the honeycomb lattice that interact via
nearest-neighbor interactions. This interaction can be parametrized
by the four-fermi coupling $\bar g$ if the four components of the
Dirac spinor $\psi^a$ are associated with the electron annihilation
operators on the two sublattices and the two Dirac points of the
honeycomb system, with the ``flavor'' $a$ representing the spin
projection \cite{Herbut2006, HerbutJuricicRoy2009}. Accordingly, $\Nf
= 2$ for \mbox{spin-$\frac{1}{2}$} particles.  In $d>2$, the coupling
$\bar g$ has positive mass dimension, reflecting the model's
perturbative nonrenormalizability. However, it is now well accepted
that the existence of an UV stable fixed point ensures that in $2<d<4$
the model is renormalizable
\textit{nonperturbatively}~\cite{RosensteinWarrPark1989}---a fact that
can be reinterpreted as maybe the simplest example for Weinberg's
asymptotic safety scenario \cite{BraunGiesScherer2011}. In the
honeycomb-lattice system the Gross-Neveu fixed point governs the phase
transition into the charge density wave phase that is expected for
large nearest-neighbor interaction~\cite{Herbut2006}. The
one-dimensional theory space defined by Eq.~\eqref{eq:action-GN} is
closed under the RG probably to any order~\cite{Gracey1991,
  VasilievDerkachovKivelStepanenko1993}, i.e., once we start with a
microscopic action of the form \eqref{eq:action-GN} no new
interactions will be generated by RG transformations. This fact has
been used to construct the fixed-point potential of the Gross-Neveu
model at arbitrary order in the fermionic field
\cite{JakovacPatkos2013}.  However, as we shall see, the RG closedness
of this model is a special property of the simple scalar-type
interaction, and does generically not hold in systems with more
complex interactions, such as in the Thirring model with a vector-type
interaction~\cite{GiesJanssen2010}. We will also see that although the
Gross-Neveu action defines a RG invariant subspace of theory space,
infinitesimal perturbations on the microscopic level may, depending on
the number of flavors $\Nf$, be RG relevant and drive the system away
from the simple Gross-Neveu theory.

In order to investigate the decomposition of the microscopic theories into universality classes---beyond just relying on the universality hypothesis---and to study the stability of the models with respect to perturbations, it is therefore mandatory to incorporate the RG flow of \textit{all} operators that are invariant under the given symmetries. This defines our $\mathrm U(\Nf)\times \mathrm U(\Nf)$ theory space: the space of all fermion theories which enjoy (at least) the symmetries of the Gross-Neveu model [Eq.~\eqref{eq:action-GN}]. These are the following:

\paragraph*{Relativistic invariance:} 
\begin{align} \label{eq:Lorentz}
\psi(x) & \mapsto e^{-\frac{i}{4} \epsilon_{\mu\nu}\gamma_{\mu\nu}}\psi(x'), &
\bar\psi(x) & \mapsto \bar\psi(x') e^{\frac{i}{4}\epsilon_{\mu\nu}\gamma_{\mu\nu}},
\end{align}
where $\gamma_{\mu\nu} = \frac{i}{2} [\gamma_\mu,\gamma_\nu]$ and
$\epsilon_{\mu\nu}$ is an antisymmetric tensor defining the rotation
axis and angle in (2+1)-dimensional space-time. Here and in the
following, where unambiguous, we suppress the 
flavor index $a$. $x'_\mu = \Lambda_{\mu\nu}^{-1} x_\nu$ with
$\Lambda_{\mu\nu}$ being the generator of the space-time rotations.
While the honeycomb lattice explicitly breaks Lorentz symmetry, it is,
however, expected to dynamically emerge close to one of the critical
points in graphene~\cite{Herbut2006,
  HerbutJuricicRoy2009}. 
Incidentally, the Lorentz
transformation of \Eqref{eq:Lorentz} can be understood as a
particular subgroup of the global spin-base transformations
$\text{SL}(4,\mathbbm{C})$ \cite{Gies:2013noa} constituted by the
similarity transformations of the Clifford algebra \cite{Pauli:1936}.

\paragraph*{Flavor symmetry:}
\begin{align}
\psi^a & \mapsto U^{ab} \psi^b, &
\bar\psi^a & \mapsto \bar\psi^b (U^\dagger)^{ba},
\end{align}
with the unitary matrix $U \in \mathrm U(\Nf)$. Flavor symmetry is generated by the (generalized) $\Nf \times \Nf$ Gell-Mann matrices $\lambda_i$, $i=1,\dots,\Nf^2-1$, together with the identity $\lambda_0 \equiv \mathbbm 1_{\Nf}$.

\paragraph*{Continuous chiral symmetry:} 
Due to our four-dimensional reducible representation of the Clifford
algebra there now exist two additional Dirac matrices, which
anticommute with all three $\gamma_\mu$: $\gamma_3$ and
$\gamma_5$. Their Hermitean product $\gamma_{35} \coloneqq i \gamma_3
\gamma_5$ generates the continuous chiral symmetry\footnote{Our
    Dirac matrix conventions are identical to those of, e.g.,
    \cite{JanssenGies2012,GiesJanssen2010} with the identification,
    i.e., simple renaming, $\gamma_3\leftrightarrow \gamma_4$ and
    $\gamma_{35}\leftrightarrow \gamma_{45}$.}
\begin{align} \label{eq:continuous-chiral}
\psi^a & \mapsto e^{i \theta \gamma_{35}} \psi^a, & \bar\psi^a &
\mapsto \bar\psi^a e^{-i \theta \gamma_{35}},
\end{align}
where $\theta \equiv \theta(a)$ may depend on the flavor index
$a=1,\dots,\Nf$. Continuous chiral symmetry and flavor symmetry
together are therefore generated by the $2\Nf^2$ matrices $\{
\lambda_0,\dots,\lambda_{\Nf^2-1} \} \otimes \{ \mathbbm 1_4,
\gamma_{35}\}$, forming the global $\mathrm U(\Nf) \times \mathrm
U(\Nf)$ symmetry of the Gross-Neveu model in the reducible
representation of the Clifford algebra. There is an alternative way to
understand this symmetry~\cite{JanssenHerbut2014}: By making use of
the orthogonal projectors $P_\mathrm{L/R} \coloneqq
\frac{1}{2}(\mathbbm 1 \pm \gamma_{35})$ with $P_\mathrm{L/R}^2 =
P_\mathrm{L/R}$, $P_\mathrm{L} P_\mathrm{R} = 0$, and $P_\mathrm{L} +
P_\mathrm{R} = \mathbbm 1$, we may decompose the four-component spinor
$\psi$ into left- and right-handed Weyl spinors $\psi_\mathrm{L/R}
\coloneqq P_\mathrm{L/R} \psi$ and $\bar\psi_\mathrm{L/R} \coloneqq
\bar\psi P_\mathrm{L/R}$, each now representing two fermionic degrees
of freedom.  [On the honeycomb lattice $\psi_\mathrm{L}$
  ($\psi_\mathrm{R}$) represents the quasiparticle excitations near
  the left (right) Dirac cone.]  The Gross-Neveu action then
decomposes into the two independent parts
\begin{align} \label{eq:action-GN-chiral}
S_\mathrm{GN} = \int d^d x \left[ \bar \psi_\mathrm{L}^a i \gamma_\mu \partial_\mu \psi_\mathrm{L}^a + \frac{\bar g}{2\Nf} \left(\bar\psi_\mathrm L^a \psi_\mathrm L^a\right)^2 \right] + (\mathrm L \leftrightarrow \mathrm R),
\end{align}
without any mixing between left- and right-handed spinors. (Note that, in contrast to the usual definition of the Weyl spinors in four dimensions, the chiral projector $P_\mathrm{L/R}$ commutes with $\gamma_0$, such that $\bar\psi_\mathrm L \psi_\mathrm R \equiv 0$.)
Eq.~\eqref{eq:action-GN-chiral} is evidently invariant under flavor rotations of the Weyl spinors:
\begin{align} \label{eq:flavor-weyl-1}
\psi_\mathrm L^a & \mapsto U_\mathrm L^{ab} \psi_\mathrm L^b, &
\bar\psi_\mathrm L^a & \mapsto \bar\psi_\mathrm L^b (U_\mathrm L^\dagger)^{ba}, \\ \label{eq:flavor-weyl-2}
\psi_\mathrm R^a & \mapsto U_\mathrm R^{ab} \psi_\mathrm R^b, &
\bar\psi_\mathrm R^a & \mapsto \bar\psi_\mathrm R^b (U_\mathrm R^\dagger)^{ba},
\end{align}
where the two unitary matrices $U_\mathrm{L/R} \in \mathrm U(\Nf)$ may be chosen independently. This constitutes the Weyl representation of the $\mathrm U(\Nf) \times \mathrm U(\Nf)$ symmetry. By separating the trace and the traceless part of the symmetry generator, we may as well split the symmetry group as
\begin{align}
\mathrm U(\Nf) \times \mathrm U(\Nf) \simeq \mathrm{SU}(\Nf) \times \mathrm{SU}(\Nf) \times \mathrm U(1)_\mathrm V \times \mathrm U(1)_\mathrm A,
\end{align}
where the phase rotations $\mathrm U(1)_\mathrm V$ (axial
transformations $\mathrm U(1)_\mathrm A$) correspond to the
transformation~\eqref{eq:flavor-weyl-1}--\eqref{eq:flavor-weyl-2} with
scalar matrices $U_\mathrm{L/R} = e^{i \theta} \mathbbm 1$
($U_\mathrm{L/R} = e^{\pm i \theta} \mathbbm 1$); the
$\mathrm{SU}(\Nf)$ factors are the remaining transformations with
traceless generators.  For the graphene system with $\Nf=2$, these
factors have the following physical meaning: $\mathrm U(1)_\mathrm V$
corresponds to charge conservation, $\mathrm U(1)_\mathrm A$ denotes
the translational symmetry on the honeycomb lattice
\cite{HerbutJuricicRoy2009}, and the two $\mathrm{SU}(2)$ factors
correspond to independent spin rotations in the two Dirac-cone sectors
\cite{JanssenHerbut2014}. While the two $\mathrm U(1)$ symmetries are
expected to hold in an effective low-energy theory of the honeycomb
lattice system, the latter may possibly (for large coupling)
not. Microscopically, only the single $\mathrm{SU}(2)$ transformation
rotating the spin simultaneously in both sectors is a
symmetry. The second $\mathrm{SU}(2)$ factor is, however, believed
to be emergent if the on-site interaction does not become too large
\cite{Herbut2006, JanssenHerbut2014}.

\paragraph*{$\mathbbm Z_2$ chiral symmetry:} 
\begin{align} \label{eq:discrete-chiral}
\psi & \mapsto \gamma_5 \psi, &
\bar\psi & \mapsto - \bar\psi \gamma_5.
\end{align}
The analogous discrete chiral symmetry with $\gamma_3$ instead of $\gamma_5$ can be obtained by combining \eqref{eq:discrete-chiral} with a prior chiral transformation \eqref{eq:continuous-chiral} with fixed $\theta =  \pi/2$ \cite{SchererGies2012}.

\paragraph*{Parity symmetry:}
We define parity transformation by inverting \textit{one} spatial coordinate, $x = (x_0,x_1,x_2) \mapsto x' = (x_0,-x_1,x_2)$,
\begin{align} \label{eq:parity}
\psi(x) & \mapsto i\gamma_1\gamma_5 \psi(x'), &
\bar\psi(x) & \mapsto \bar\psi(x') i\gamma_1\gamma_5.
\end{align}
Note that different definitions of parity symmetry are in principle possible, when combining \eqref{eq:parity} with chiral transformations \eqref{eq:continuous-chiral}. On the honeycomb lattice, the above form corresponds to the reflection symmetry which exchanges the two Dirac points while not exchanging the sublattice labels \cite{HerbutJuricicRoy2009}. The second reflection symmetry of the honeycomb lattice, which exchanges the spinor components belonging to the two different sublattices, is obtained by a combination of \eqref{eq:discrete-chiral}, \eqref{eq:parity}, and the rotational symmetry from \eqref{eq:Lorentz} with $\epsilon_{\mu\nu} = \pi (\delta_{\mu 1}\delta_{\nu 2} - \delta_{\mu 2}\delta_{\nu 1})$. 

Further discrete transformations may be defined, such as time reversal and charge conjugation, both of which leave the Gross-Neveu action \eqref{eq:action-GN} invariant. For simplicity, we do not list them here, since they do not lead to any further restraints on possible operators in the theory. The $\mathrm U(\Nf) \times \mathrm U(\Nf)$ Gross-Neveu theory space is uniquely determined by the above given set of symmetry transformations.

We now classify all pointlike operators up to the four-fermion
level with respect to their symmetry. Let us start with the
two-fermion terms, which represent the building blocks of the
higher-order operators. Flavor symmetry ensures the form
\begin{align}\label{eq:two-fermion}
\bar\psi^a \mathcal O \psi^a
\end{align}
with a $4\times 4$ matrix $\mathcal O$. A basis in the 16-dimensional space of $4\times 4$ operators is given by the gamma matrices and their products:
\begin{align} 
\mathcal O \in \Span \{ \mathbbm 1_4, \gamma_\mu, \gamma_3, \gamma_5, \gamma_{\mu\nu}, \gamma_{35}, i\gamma_\mu\gamma_3, i\gamma_\mu\gamma_5\}
\end{align}
The two-fermion term \eqref{eq:two-fermion} will be invariant under the continuous chiral symmetry \eqref{eq:continuous-chiral}, if $[\mathcal O, \gamma_{35}]=0$, which restricts $\mathcal O \in \Span \{ \mathbbm 1_4, \gamma_\mu, \gamma_{\mu\nu}, \gamma_{35} \}$. For invariance under discrete chiral symmetry \eqref{eq:discrete-chiral}, however, we demand $\mathcal O$ to anticommute with $\gamma_5$, which requires $\mathcal O \in \Span \{ \gamma_\mu, \gamma_3, \gamma_{35}, i \gamma_\mu\gamma_5 \}$. Finally, parity invariance implies the commutation relation $[\mathcal O, i\gamma_1\gamma_5]=0$ and thus $\mathcal O \in \Span \{ \mathbbm 1_4, \gamma_0, \gamma_2, \gamma_3,\gamma_{02}, i\gamma_0\gamma_3,i\gamma_2\gamma_3,i\gamma_1\gamma_5\}$. If we additionally require Lorentz invariance, these restrictions mutually exclude each other, and there exists no two-fermion term that is invariant under all symmetries of the $\mathrm U(\Nf) \times \mathrm U(\Nf)$ Gross-Neveu model. If we relax the condition of Lorentz symmetry down to the invariance under just spatial rotations, the only invariant two-fermion term would be $\bar\psi \gamma_0 \psi$. In the honeycomb-lattice system this represents the quasiparticle density, which vanishes at half filling.

On the level of four-fermion terms there are in principle two different types possible,
\begin{align} \label{eq:four-fermion}
(\bar\psi^a\mathcal O \psi^a)(\bar\psi^b \mathcal Q \psi^b) \quad \text{and} \quad
(\bar\psi^a\mathcal O \psi^b)(\bar\psi^b \mathcal Q \psi^a),
\end{align}
which we will refer to as having singlet and nonsinglet flavor structure, 
respectively. Not all of these terms, however, are independent: With the help of 
Fierz identities, we can always rewrite terms of the second type (with 
nonsinglet flavor structure) as a linear combination of terms of the first type 
(with singlet flavor structure), see below. To begin with, it thus suffices to 
determine the invariant terms with singlet flavor structure; the nonsinglet-type 
ones can subsequently be obtained by Fierz identities. From the above discussion 
it is clear, that only the terms with $\mathcal O = \mathcal Q$ will have a 
chance to be invariant under the given symmetries: Each of the (anti)commutation 
conditions, which lead to restrictions of the two-fermion terms, $[\mathcal 
O,\gamma_{35}] = 0$ (continuous chiral symmetry), $\{\mathcal O, \gamma_5\} = 0$ 
(discrete chiral symmetry), and $[\mathcal O, \gamma_{15}]$ (parity symmetry), 
divides the 16-dimensional space of $4\times 4$ matrices into two equally large 
8 
dimensional subspaces of matrices which do and do not respectively fulfill the particular (anti)commutation relation. Invariance of the four-fermion term \eqref{eq:four-fermion} requires $\mathcal O$ and $\mathcal Q$ to be in the same subspace for each of the symmetries. Together with Lorentz invariance, this is not simultaneously achievable for all three above symmetries if $\mathcal O \neq \mathcal Q$. Furthermore, for the flavor singlet term in Eq.~\eqref{eq:four-fermion} to be invariant under the continuous chiral symmetry we need $\mathcal O = \mathcal Q$ to commute with $\gamma_{35}$. A basis of flavor-singlet four-fermion terms invariant under the above set of symmetries is therefore
\begin{align} \label{eq:basis-standard-A}
(S)^2 & \coloneqq (\bar\psi^a \psi^a)^2, &
(P)^2 & \coloneqq (\bar\psi^a \gamma_{35} \psi^a)^2, \\ \label{eq:basis-standard-B}
(V)^2 & \coloneqq (\bar\psi^a \gamma_{\mu} \psi^a)^2, &
(T)^2 & \coloneqq \frac{1}{2}(\bar\psi^a\gamma_{\mu\nu}\psi^a)^2.
\end{align}
Our theory space includes several previously investigated systems: 
\begin{enumerate}[(1)]
\item The Gross-Neveu model in the four-dimensional reducible representation of the Clifford algebra (``reducible Gross-Neveu model'') with Lagrangian $\mathcal L = \bar\psi i \gamma_\mu \partial_\mu \psi + \bar g_S (S)^2$ has been discussed in Ref.~\cite{BraunGiesScherer2011}. 
\item If we choose a gamma-matrix basis in which 
$\gamma_{35} = 
\left(\begin{smallmatrix}
 \mathbbm{1}_2 \\
 & -\mathbbm{1}_2
\end{smallmatrix}\right)$
the $\Nf$ four-component Dirac spinors $\psi^a$, $a=1,\dots,\Nf$, can be reduced to $2\Nf$ two-component Dirac spinors $\chi^i$, $i=1,\dots,2\Nf$ by means of
\begin{align}
\psi \equiv
\begin{pmatrix}
 \chi^a \\
 \chi^{a+\Nf}
\end{pmatrix}
\quad 
\text{and}
\quad
\bar\psi \equiv
\begin{pmatrix}
 \bar\chi^a \\
 -\bar\chi^{a+\Nf}
\end{pmatrix}.
\end{align}
$\chi$ and $\bar\chi$ are related to the Weyl spinors by 
$\psi_\mathrm{L}^a \pm \psi_\mathrm{R}^a = 
\left(\begin{smallmatrix}
 \chi^{a} \\
 \pm \chi^{a+\Nf}
\end{smallmatrix}\right)$
and
$\bar\psi_\mathrm{L}^a \pm \bar\psi_\mathrm{R}^a = 
\left(\begin{smallmatrix}
 \bar\chi^a \\
 \mp \bar\chi^{a+\Nf}
\end{smallmatrix}\right)$.
In terms of the $2\Nf$ flavors of two-component spinors $\chi^i$ the interaction channel $(P)^2$ can rewritten as a (standard) Gross-Neveu interaction~\cite{JanssenGies2012},
\begin{align} \label{eq:P2-barchi-chi}
 (P)^2 = (\bar\chi^i \chi)^2.
\end{align}
The Lagrangian $\mathcal L = \bar\psi i \gamma_\mu \partial_\mu \psi + \bar g_P (P)^2$ thus describes the Gross-Neveu model in the irreducible representation of the Clifford algebra (``irreducible Gross-Neveu model''). Its critical behavior has been studied in Refs.~\cite{RosensteinWarrPark1989, RosensteinYuKovner1993, Gracey1991, VasilievDerkachovKivelStepanenko1993, HandsKocicKogut1993, RosaVitaleWetterich2001}.
\item The system with Lagrangian $\mathcal L = \bar\psi i \gamma_\mu \partial_\mu \psi + \bar g_V (V)^2$ is known as the Thirring model and has also been subject of several previous investigations~\cite{Parisi1975, DelDebbioHandsMehegan1997, ChristofiHandsStrouthos2007, Chandrasekharan2010, ChandrasekharanLi2012-b, Mesterhazy2012, GiesJanssen2010, JanssenGies2012}.
\end{enumerate}
%
%Within the present approach that considers the whole space of theories we can thus ask questions such as how these and related systems decompose into universality classes and why.

An analogous discussion for the flavor nonsinglet terms [second term
in Eq.~\eqref{eq:four-fermion}] is possible, but can be kept
short
with the help of the Fierz identities \cite{GiesJanssen2010}:
\begin{align} \label{eq:Fierz-A}
(S)^2 & = - \frac{1}{4}\left[ (S^D)^2 + (P^D)^2 + (V^D)^2 + (A^D)^2 \right], \\
(P)^2 & = - \frac{1}{4}\left[ (S^D)^2 - (P^D)^2 + (V^D)^2 - (A^D)^2 \right], \\
(V)^2 & = - \frac{1}{4}\left[ 3(S^D)^2 -3 (P^D)^2 - (V^D)^2 + (A^D)^2 \right], \\ \label{eq:Fierz-D}
(T)^2 & = - \frac{1}{4}\left[ 3(S^D)^2 +3 (P^D)^2 - (V^D)^2 - (A^D)^2 \right],
\end{align}
where we have abbreviated
\begin{align} \label{eq:basis-dual-A}
(S^D)^2 & \coloneqq (\psi^a \psi^b)^2 + (\psi^a \gamma_{35} \psi^b)^2, \\
(P^D)^2 & \coloneqq (\psi^a \gamma_3 \psi^b)^2 + (\psi^a \gamma_5 \psi^b)^2, \\
(V^D)^2 & \coloneqq (\psi^a \gamma_\mu \psi^b)^2 + \frac{1}{2} (\psi^a \gamma_{\mu\nu} \psi^b)^2, \\ \label{eq:basis-dual-D}
(A^D)^2 & \coloneqq (\psi^a i\gamma_\mu \gamma_3 \psi^b)^2 + (\psi^a i\gamma_\mu \gamma_5 \psi^b)^2.
\end{align}
This yields an (invertible) one-to-one correspondence between vectors in the space of flavor singlet terms with their ``dual'' counterparts in the space of flavor nonsinglet terms. Therefore, any invariant flavor nonsinglet term must be a linear combination of the dual basis vectors $(S^D)^2$, $(P^D)^2$, $(V^D)^2$, and $(A^D)^2$. A full basis of fermionic four-point functions in the limit of pointlike interactions is thus given by, e.g., the four terms in Eqs.~\eqref{eq:basis-standard-A}--\eqref{eq:basis-standard-B} or the four terms in Eqs.~\eqref{eq:basis-dual-A}--\eqref{eq:basis-dual-D}, or a combination thereof. Put differently, any four-fermion theory with the symmetries of the $\mathrm U(\Nf) \times \mathrm U(\Nf)$ Gross-Neveu model represents one point in the theory space spanned by a set of four basis vectors from Eqs.~\eqref{eq:basis-standard-A}--\eqref{eq:basis-standard-B}, \eqref{eq:basis-dual-A}--\eqref{eq:basis-dual-D}. In the following, we will investigate the RG evolution in this theory space 
with a particular focus on possible fixed points.

Before proceeding, however, let us make a comment on the special case of single fermion flavor ($\Nf = 1$), which in the condensed-matter applications corresponds to the common simplified model of spinless electrons. In this case, we have two further relations between the interaction terms
\begin{align}
(S^D)^2 & = (\bar\psi \psi)^2 + (\bar\psi \gamma_{35}\psi)^2 = (S)^2+(P)^2, \label{eq:Nf1-SD2}\\
(V^D)^2 & = (\bar\psi\gamma_\mu\psi)^2 + \frac{1}{2}(\bar\psi\gamma_{\mu\nu}\psi)^2 = (V)^2+(T)^2, \label{eq:Nf1-VD2}
\end{align}
in addition to the Fierz identities.  Evaluation of the corresponding
matrix rank yet shows that only one of these two is independent and,
e.g., Eq.~\eqref{eq:Nf1-SD2} can be obtained by linear combination of
Eq.~\eqref{eq:Nf1-VD2} with the Fierz identities
\eqref{eq:Fierz-A}--\eqref{eq:Fierz-D}. Thus, for $\Nf=1$ a
``Fierz-complete'' basis is given by just three independent
interaction terms, e.g., the flavor singlets $(S)^2$, $(P)^2$, and
$(V)^2$. The corresponding relation among the singlet invariants is
\begin{equation}
(T)^2 = - 3(S)^2 - 3(P)^2 - (V)^2, \quad \text{for}\,\, \Nf=1.
\label{eq:FierzNf1}
\end{equation}
This is in agreement with the previous study of spinless
fermions on the honeycomb lattice \cite{HerbutJuricicRoy2009}.

%%%%%%%%%%%%%%%%%%%%%%%%%%%%%%%%%%%%%%%%%%%%%%%%%%%%%%%%%%%%%%%%%%%%%%%%%%%%%%%%
\section{Functional Renormalization Group}
\label{sec:frg}
In order to compute the RG beta functions, we use the functional
renormalization group in terms of Wetterich's evolution equation for
the effective average action $\Gamma_k$ \cite{Wetterich1993},
\begin{align} \label{eq:Wetterich}
\partial_k \Gamma_k & = \frac{1}{2} \STr\left[ \partial_k R_k(\Gamma_k^{(2)} + R_k)^{-1}\right],
\end{align}
where $\Gamma_k$ denotes a scale-dependent effective action as a
function of an infrared RG scale scale $k \in [0,\Lambda]$ ($\Lambda$
being the UV cutoff). This action interpolates between the microscopic
action $S$ for $k \to \Lambda$, and the full quantum effective action
$\Gamma$ (generating functional of one-particle irreducible Green's
functions) for $k \to 0$. The quantity $\Gamma_k^{(2)}$ denotes the
corresponding Hessian of the effective action, and the function
regulator $R_k$ defines the details of the regularization
procedure. Within the approximation used in this work, all our results
given below turn out to be independent of the regularization scheme.
For reviews on the functional RG, see
Refs.~\cite{BergesTetradisWetterich2002, KopietzBartoschSchuetz2010,
  MetznerSalmhoferHonerkampMedenSchoenhammer2012}. 

While the Wetterich equation \eqref{eq:Wetterich} is an exact identity
for the effective average action, it is difficult
to solve it without the use of suitable approximation schemes. The
four-fermion theories considered in this work exhibit a lower critical
dimension of $d=2$, at which the four-fermion couplings become
marginal. In dimension $d=2+\epsilon$ for small $\epsilon$ any
interacting fixed point will therefore be of the order $g_* = \mathcal
O(\epsilon)$ and thus accessible via a (renormalized) perturbative
expansion. An analogous argument can be made for a large number of
fermion flavors $\Nf$. Any higher-order fermionic term, as well as
momentum-dependent fermionic vertices will then be irrelevant at such
an interacting fixed point, and the effective average action in the
vicinity of the fixed point will have the truncated form
\begin{multline} \label{eq:truncation}
\Gamma_k = \int d^dx \left[ \bar \psi^a i\gamma_\mu \partial_\mu \psi^a + \frac{\bar g_{S,k}}{2\Nf} (S)^2 +  \frac{\bar g_{P,k}}{2\Nf} (P)^2 \right. \\
 \left. + \frac{\bar g_{V,k}}{2\Nf} (V)^2 + \frac{\bar g_{T,k}}{2\Nf} (T)^2 \right],
\end{multline}
where we have used the flavor-singlet basis from
Eqs.~\eqref{eq:basis-standard-A}--\eqref{eq:basis-standard-B} with
scale-dependent four-fermion couplings $\bar g_{S,k}$, $\bar g_{P,k}$,
$\bar g_{V,k}$, and $\bar g_{T,k}$.  For $\Nf=1$, the last term
$\propto (T)^2$ in Eq.~\eqref{eq:truncation} can be rewritten as a
linear combination of the three former terms $(S)^2$, $(P)^2$, and
$(V)^2$, and $\Gamma_k$ would be spanned by just three couplings.  We
have neglected the possibility of a wave-function renormalization for
the fermions, as the fermion anomalous dimension is known to vanish in
the limit of pointlike fermionic interactions \cite{GiesJanssen2010,
  JakovacPatkos2013}. In the following, we will use
Eq.~\eqref{eq:truncation} as an ansatz for $\Gamma_k$ to solve the
Wetterich equation \eqref{eq:Wetterich} approximately. As all omitted
terms are perturbatively RG irrelevant, this truncation will become
exact to first order in $\epsilon$ close to the lower critical
dimension, as well as for large number of flavors $\Nf$. For the
physical cases with $\epsilon = 1$ and small $\Nf$,
Eq.~\eqref{eq:truncation} can of course only be viewed as the simplest
possible truncation within, e.g., a systematic expansion of $\Gamma_k$
in terms of derivatives. Beyond the perturbative domain, higher-order
terms (e.g., momentum-dependent terms) would in principle have to be
taken into account, and the stability of our results against the
inclusion of such terms would have to be verified. In the framework of
the functional RG this can be done, e.g., by means of partial
\cite{LeeStrackSachdev2013, JanssenHerbut2014} or dynamical
\cite{GiesWetterich2002, Pawlowski2007, JanssenGies2012,Braun:2014ata}
bosonization techniques, or by various decomposition schemes in
momentum space \cite{ MetznerSalmhoferHonerkampMedenSchoenhammer2012,
  GieringSalmhofer2012}. A third option is offered by working
with full potentials for fermion bilinears and determining the
solution of the flow on all scales on a larger function space
including {\em weak solutions} \cite{Aoki:2014ola}, as has been used
for 3d models in \cite{JakovacPatkos2013}. This will be left
for future work, and we will here confine ourselves to the study of
the RG flow in the limit of pointlike (momentum-independent)
four-fermion couplings.

%%%%%%%%%%%%%%%%%%%%%%%%%%%%%%%%%%%%%%%%%%%%%%%%%%%%%%%%%%%%%%%%%%%%%%%%%%%%%%%%
\section{General formula for 4-fermi beta functions}
\label{sec:general-4-fermi}

By plugging the ansatz for $\Gamma_k$ [Eq.~\eqref{eq:truncation}] into
the Wetterich equation \eqref{eq:Wetterich}, the flow equations for
the four-fermion couplings can straightforwardly (though possibly
somewhat tediously) be obtained by equating the coefficients on the
left and right hand side of the equation. However, it appears
worthwhile to cast Eq.~\eqref{eq:truncation} into a more general form
\begin{align} \label{eq:truncation-general}
\Gamma_k & = \int d^dx \left[ \bar \psi^a i\gamma_\mu \partial_\mu \psi^a +
\sum_i \frac{\bar g_{i,k}}{2\Nf} (\bar\psi^a \mathcal O_i \psi^a)^2 \right],
\end{align}
describing a general massless fermion system interacting via several
short-range interactions. Let us assume that the interactions in
Eq.~\eqref{eq:truncation-general} represent a ``full basis'' of
fermionic four-point functions in the pointlike limit in the sense of
Sec.~\ref{sec:symmetry}, thus including all flavor-singlet terms which
are invariant under a given symmetry. Invariant flavor-nonsinglet
terms can be rewritten as a linear combination of the above by means
of Fierz identities, and in that sense the basis is ``Fierz
complete''.  As we shall see in this section, the functional RG
equation will allow us to compute the beta functions for this general
ansatz of four-fermion interactions.  The resulting formula is readily
automatable within a computer algebra system. This will allow to
compute the one-loop beta functions in various fermion systems in
arbitrary dimension. E.g., it should be applicable to derive effective
theories for several condensed-matter systems featuring chiral
fermions, such as single- \cite{HerbutJuricicRoy2009} or bilayer
\cite{Vafek2010} graphene, or in the (3+1)-dimensional systems with
linear \cite{VafekVishwanath2014} or quadratic
\cite{HerbutJanssen2014} dispersion relation---systems in which
generically a large number of interactions are compatible with the
given microscopic symmetries.%
\footnote{We note that while we here have assumed relativistic fermions, the crucial point for the derivation of the beta functions is the fermion 
propagator's Dirac-matrix structure $\sim \gamma_\mu$, which has an analogous form in nonrelativistic chiral fermion systems with quadratic dispersion relation both in 2+1 \cite{SunYaoFradkinKivelson2009} as well as 3+1 dimensions \cite{HerbutJanssen2014}. We believe that an analogous formula as derived in this section should also be possible to derive for these systems.}

In order to formalize the procedure of equating coefficients, we introduce the ``projector''
\begin{multline}
\hat P_{i} \coloneqq \sum_j \sum_{a,b} \sum_{\alpha,\beta,\gamma,\delta} c_{ij}
\mathcal O_{j,\gamma \beta} \mathcal O_{j,\delta\alpha}  \\
\times \left. \frac{\overrightarrow\delta}{\delta \bar\psi_\alpha^a}\frac{\overrightarrow\delta}{\delta \bar\psi_\beta^b} (\ \cdot \ ) \frac{\overleftarrow\delta}{\delta \psi_\gamma^b}\frac{\overleftarrow\delta}{\delta \psi_\delta^a}\right|_{\bar\psi=\psi=0}
\end{multline}
with to-be-defined coefficients $c_{ij}$. In the functional derivatives, the Greek letters $\alpha, \beta, \gamma, \delta =1,\dots, d_\gamma$ refer to the spinor index, while Latin letters $a,b = 1,\dots, \Nf$ denote the flavor index. $d_\gamma$ is the dimension of the representation of the Clifford algebra, $\{\gamma_\mu,\gamma_\nu\} = 2\delta_{\mu\nu} \mathbbm 1_{d_\gamma}$.
When applied to our ansatz for the effective action [Eq.~\eqref{eq:truncation-general}], $\hat P_i$ gives
\begin{align}
\hat P_{i} \Gamma = \sum_{j,l} c_{ij} M_{jl} \bar g_{l} 
\end{align}
with the symmetric matrix 
\begin{align}
M_{jl} = \Nf \left[\Tr(\mathcal O_j \mathcal O_l)\right]^2  - \Tr (\mathcal O_j
\mathcal O_l \mathcal O_j \mathcal O_l) \qquad \text{(no sum)}.
\end{align}
(Wherever possible, we suppress the scale index $k$ of the
coupling $\bar g_l \equiv \bar g_{l,k}$ in the following to avoid confusion.)
If we choose the
coefficients $c_{ij}$ such that
\begin{align}
\sum_{j} c_{ij} M_{jl} = \delta_{il} \qquad \Rightarrow \qquad c_{ij} = (M^{-1})_{ij},
%\left[ \Nf \Tr^2(\mathcal O^j \mathcal O^l)  - \Tr (\mathcal O^j \mathcal O^l \mathcal O^j \mathcal O^l) \right] = 
\end{align}
we find that by applying $\hat P_i$ on the Wetterich equation we ``extract'' the beta function for the $i$th coupling constant,
\begin{align} 
\partial_k \bar g_{i} & = \hat P_i (\partial_k\Gamma_k) \nonumber
\\ \label{eq:supertrace}
 & = \frac{1}{2} \hat P_i\STr\left[ \partial_k R_k(\Gamma_k^{(2)} + R_k)^{-1}\right].
\end{align}
This can be further simplified by introducing the scale-derivative $\tilde
\partial_k \coloneqq (\partial_k R_k) \frac{\delta}{\delta R_k}$, which acts
only on the regulator's $k$-dependence~\cite{BergesTetradisWetterich2002},
\begin{align} \label{eq:logarithm}
\partial_k R_k(\Gamma_k^{(2)} + R_k)^{-1} = \tilde \partial_k \ln (\Gamma_k^{(2)} + R_k),
\end{align}
and expanding the logarithm in powers of interactions
\begin{multline} \label{eq:logarithm-series}
\ln (\Gamma_k^{(2)} + R_k) = \ln (\Gamma_{k,0}^{(2)} + R_k) \\
 + \sum_{n=1}^\infty \frac{(-1)^{n+1}}{n} \left[\Delta \Gamma_k^{(2)} (\Gamma_{k,0} + R_k)^{-1}\right]^n.
\end{multline}
Here we have split the fluctuation matrix into its
field-independent propagator part $\Gamma^{(2)}_{k,0} = \Gamma^{(2)}_k
|_{\bar\psi=\psi=0}$ (which is straightforwardly invertible) and the
fluctuation part $\Delta\Gamma_k^{(2)} = \Gamma^{(2)}_k -
\Gamma^{(2)}_{k,0}$. For the four-fermion theories,
$\Delta\Gamma^{(2)}_k$ is quadratic in $\psi, \bar\psi$, and thus the
series in Eq.~\eqref{eq:logarithm-series} terminates after $n = 2$
when applying the ``projector'' $\hat P_i$. In massless systems also
the leading term for $n=1$ vanishes for reasons of symmetry. The only
nonvanishing term when plugging Eqs.~\eqref{eq:logarithm} and
\eqref{eq:logarithm-series} into \eqref{eq:supertrace} is then
the quadratic term,
\begin{align} \label{eq:flow-eq-second-order}
\partial_k \bar g_{i} = -\frac{1}{4}\hat P_i \tilde \partial_k \STr
\left[\Delta \Gamma_k^{(2)} (\Gamma_{k,0} + R_k)^{-1}\right]^2,
\end{align}
ensuring that the flow of the four-fermi coupling has the usual (one-loop) quadratic form
\begin{align}
\partial_k \bar g_{i} = \frac{4 v_d}{\Nf k^{d-1}} \ell_{1}^{(\mathrm F)d}
\sum_{j,l} \bar g_j A_{i,jl} \bar g_l,
\end{align}
with the coefficient matrix $A_{i,jl}$ to be determined.  Here, we
have used the standard abbreviations $v_d =
\frac{1}{4}\text{vol}(S^{d-1})/(2\pi)^d = [2^{d+1}\pi^{d/2}
  \Gamma(d/2)]^{-1}$, which arises from the angular part of the loop
integral, and the dimensionless threshold function $\ell_1^{(\mathrm
  F)d}$, representing the loop integral's radial part. It incorporates
the regulator dependence and can be defined by \cite{GiesJanssen2010}
\begin{align}
\ell_1^{(\mathrm F)d} = - \partial_k \int_0^{\Lambda^2} dq^2 \frac{q^{d-4}}{\left[1+r(q^2/k^2)\right]^2}.
\end{align}
In terms of rescaled dimensionless coupling constants \mbox{$g_i \coloneqq 4
v_d \ell_1^{(\mathrm F)d} k^{d-2} \bar g_{i}$} the beta function reads
\begin{align} \label{eq:beta-general}
\partial_t g_i = (d-2) g_i + \frac{1}{\Nf} \sum_{j,l} g_j A_{i,jl}g_l, \qquad
t = \ln (k/\Lambda).
\end{align}
By employing the ansatz \eqref{eq:truncation-general} for $\Gamma_k$, Eq.~\eqref{eq:flow-eq-second-order} gives after some elementary algebra the coefficient matrix
\begin{multline}
\label{eq:coefficients}
 A_{i,jl}  = 
\frac{1}{d} \sum_{m} c_{im} \Bigl\{ \Tr(\mathcal O_{l} \gamma_\mu \mathcal O_{j} \mathcal O_{m} \mathcal O_{l} \gamma_\mu \mathcal O_{j} \mathcal O_{m}) \\
 - \Tr(\mathcal O_{l} \gamma_\mu \mathcal O_{j} \mathcal O_{m} \mathcal O_{j} \gamma_\mu \mathcal O_{l} \mathcal O_{m})  \\
  - \Tr(\mathcal O_{j} \mathcal O_{m} \mathcal O_{l} \gamma_\mu \mathcal O_{j} \gamma_\mu \mathcal O_{l} \mathcal O_{m}) \\
 - \Tr(\mathcal O_{l} \mathcal O_{m} \mathcal O_{j} \gamma_\mu \mathcal O_{l} \gamma_\mu \mathcal O_{j} \mathcal O_{m}) \\
  + \Nf \bigl[ - \Tr(\mathcal O_{l} \gamma_\mu \mathcal O_{j} \mathcal O_{m}) \Tr(\mathcal O_{l} \gamma_\mu \mathcal O_{j} \mathcal O_{m}) \\
 + \Tr(\mathcal O_{l} \gamma_\mu \mathcal O_{j} \mathcal O_{m})\Tr(\mathcal O_{j} \gamma_\mu \mathcal O_{l} \mathcal O_{m}) \\
 + \Tr(\mathcal O_{j} \mathcal O_{m} \mathcal O_{l} \mathcal O_{m})\Tr(\gamma_\mu \mathcal O_{j} \gamma_\mu \mathcal O_{l}) \\
 + \Tr(\mathcal O_{j} \mathcal O_{m})\Tr(\mathcal O_{l} \gamma_\mu \mathcal O_{j} \gamma_\mu \mathcal O_{l} \mathcal O_{m}) \\
 + \Tr(\mathcal O_{l} \mathcal O_{m})\Tr(\mathcal O_{j} \gamma_\mu \mathcal O_{l} \gamma_\mu \mathcal O_{j} \mathcal O_{m})\bigr]  \\
  - \Nf^2 \Tr(\mathcal O_{j} \mathcal O_{m})\Tr(\mathcal O_{l} \mathcal O_{m})\Tr(\gamma_\mu \mathcal O_{j} \gamma_\mu \mathcal O_{l})\Bigr\}.
\end{multline}
Eqs.~\eqref{eq:beta-general}--\eqref{eq:coefficients} represent the main result of this section. They are (up to the rescaling) exactly the one-loop beta functions as they would have been obtained within the standard Wilsonian momentum-shell RG approach. In particular due to its simple possible implementation within a computer algebra system, we expect this general formula to be valuable also beyond the scope of the models considered in this work. We will use it in the following to compute the flow of the four couplings present in the $\mathrm U(\Nf) \times \mathrm U(\Nf)$ Gross-Neveu theory space [Eq.~\eqref{eq:truncation}].

%%%%%%%%%%%%%%%%%%%%%%%%%%%%%%%%%%%%%%%%%%%%%%%%%%%%%%%%%%%%%%%%%%%%%%%%%%%%%%%%
\section{Flow equations}
\label{sec:flow-equations}
A straightforward evaluation of the traces in Eq.~\eqref{eq:coefficients} gives the beta functions for the pointlike four-fermion couplings in the four-dimensional Gross-Neveu theory space for $\Nf>1$
\begin{align}
\partial_t g_{S} & = (d-2)g_{S} + \frac{1}{\Nf} \bigl[(-4 \Nf + 2) g_{S}^2 +
 \nonumber \\ & \quad
+ g_{S}(2 g_{P}+6 g_{V} + 6 g_{T}) + 8 g_{V} g_{T} \bigr], \label{eq:flow-A}  \allowdisplaybreaks[1] \\
\partial_t g_{P} & = (d-2)g_{P} + \frac{1}{\Nf} \bigl[(-4 \Nf + 2) g_{P}^2 + 
 \nonumber \\ & \quad
+g_{P}(2 g_{S}+6 g_{V} + 6 g_{T}) + 4 g_{V}^2 + 4 g_{T}^2\bigr], \allowdisplaybreaks[2] \\
\partial_t g_{V} & = (d-2)g_{V} + \frac{1}{\Nf} \biggl[\frac{4 \Nf + 2}{3} g_{V}^2 + 
 \nonumber \\ & \quad
+g_{V}\left(-\frac{2}{3} g_{S}+2 g_{P} + \frac{2}{3} g_{T}\right) + \frac{8}{3} g_{S} g_{T} \biggr],  \allowdisplaybreaks[1] \\
\partial_t g_{T} & = (d-2)g_{T} + \frac{1}{\Nf} \biggl[\frac{4 \Nf + 2}{3} g_{T}^2 + 
 \nonumber \\ & \quad
+g_{T}\left(-\frac{2}{3} g_{S}+2 g_{P} + \frac{2}{3} g_{V}\right) + \frac{8}{3} g_{S} g_{V} \biggr]. \label{eq:flow-D}
\end{align}
We observe that these flow equations are symmetric under the exchange
of $g_V \leftrightarrow g_T$.  They generalize several previous RG
approaches to relativistic fermion systems:

Setting $g_P = g_V = g_T \equiv 0$ defines the $\mathrm U(\Nf) \times \mathrm U(\Nf)$ Gross-Neveu model with four-component Dirac fermions. The remaining flow equation 
\begin{align}
\partial_t g_S = (d-2) g_S - \frac{4\Nf-2}{\Nf} g_S^2
\end{align} 
agrees with the previous calculation~\cite{BraunGiesScherer2011}. We
also see that this subspace is closed under the RG, and $g_P$, $g_V$,
and $g_T$ will not be generated if they all vanish at the initial
scale.

The subspace $g_S = g_T \equiv 0$ has also been considered earlier,
and our flow equations agree with the former
work~\cite{GiesJanssen2010}. In this subspace the system's $\mathbbm
Z_2 \times \mathrm U(\Nf) \times \mathrm U(\Nf)$ chiral symmetry is
elevated to $\mathrm U(2\Nf)$, generated by the $4\Nf^2$ matrices
$\{\lambda_0,\dots,\lambda_{\Nf^2-1}\} \otimes \{\mathbbm 1_4,
\gamma_{3}, \gamma_{5}, \gamma_{35}\}$, and RG closedness is protected
by symmetry. This is the theory space of the Thirring model in $2<d<4$
dimensions \cite{JanssenGies2012, GiesJanssen2010}. In the following,
we will refer to it as $\mathrm U(2\Nf)$ or Thirring subspace.

For $\Nf=1$, when the theory space is just three-dimensional, one of
the four couplings in Eqs.~\eqref{eq:flow-A}--\eqref{eq:flow-D} can be
completely eliminated. Using \Eqref{eq:FierzNf1}, we find
\begin{align}
& g_{S} (S)^2+ g_{P} (P)^2+ g_{V} (V)^2+ g_{T} (T)^2 
=\nonumber \\
&  \quad (g_{S} -3 g_{T}) (S)^2 + (g_{P}-3 g_{T}) (P)^2+
(g_{V}- g_{T}) (V)^2.
\end{align}
If we shift the couplings as
\begin{align}
g_S - 3 g_T &\mapsto g_S, &
g_P - 3 g_T &\mapsto g_P, &
g_V - g_T &\mapsto g_V,
\end{align}
the corresponding shifted beta functions for $\Nf=1$ then indeed become independent of the fourth coupling $g_T$,
\begin{align}
\partial_t g_{S} & = (d-2)g_{S} + 2 g_{S}(-g_{S} + g_{P}- g_{V}) \label{eq:flow-Nf1-A},\\
\partial_t g_{P} & = (d-2)g_{P} + 2 g_{P}(- g_{P} + g_{S}+ 3 g_{V}) \nonumber \\
&\quad + 4 g_{V}(g_{V}-2g_{S}),\\
\partial_t g_{V} & = (d-2)g_{V} + 2 g_{V}\left(g_{V}-\tfrac{5}{3} g_{S} + g_{P}\right).  \label{eq:flow-Nf1-C}
\end{align}
We note that Eqs.~\eqref{eq:flow-Nf1-A}--\eqref{eq:flow-Nf1-C} are equivalent to the previous one-loop flow equations for the spinless-fermion system on the honeycomb lattice \cite{HerbutJuricicRoy2009}.%
\footnote{
The equivalence can be readily seen by using the coupling transformation $g_\alpha = - g_V$, $g_{C1} = g_S - g_V$, and $g_{D2} = g_P - 2 g_V$, where $g_{\alpha,C1,D2}$ are the three couplings used in Ref.~\cite{HerbutJuricicRoy2009}, assuming the full Lorentz symmetry.
} 
The flow equations thus again reflect the fact that for $\Nf=1$ different values for the coupling $g_T$ (in terms of the shifted couplings) do not correspond to different physical points in theory space, but rather can be mapped onto each other by means of the Fierz identities; the linearly dependent set of the four elements $(S)^2$, $(P)^2$, $(V)^2$, and $(T)^2$, which span the theory space, is no longer a basis for $\Nf=1$, but an overcomplete \textit{frame}, as used, e.g., in the theory of signal processing \cite{Mallat2008}.
Of course, the same would happen for $\Nf>1$ if further linearly dependent (e.g., flavor-nonsinglet) terms were to be added to the effective action. 

Let us express a warning at this point: When counting the number of
(physically distinct) fixed points, and the number of their
accompanying relevant and irrelevant directions (as we shall do in the
following) it is thus important to take all identities between the
fermionic terms into account and reduce the frame to a (linearly
independent) \textit{basis}: Otherwise, e.g., critical fixed points,
having a single relevant direction in the irreducible basis, may
develop spurious additional relevant (or irrelevant) directions in the
redundant directions, preventing them to be identified as critical
fixed points in the overcomplete frame. On the other hand, if by means
of a single- or few-channel approximation we \textit{a priori} neglect
particular (physical) directions in theory space, which in principle
would be compatible with the symmetry, a fixed point with several RG
relevant directions could falsely appear as critical fixed point in
such a truncated description. In this work, we therefore advocate the
use of a Fierz-complete irreducible basis of theory space in order to
find the number of fixed points and their relevant and irrelevant
directions.

%%%%%%%%%%%%%%%%%%%%%%%%%%%%%%%%%%%%%%%%%%%%%%%%%%%%%%%%%%%%%%%%%%%%%%%%%%%%%%%%
%\section{UV fixed-point structure}
%
\section{General properties of the one-loop flow}
\label{sec:general-one-loop}
The topology of the flow is determined by the fixed points $\boldsymbol{g}^*$ of the RG, where all beta functions vanish, $\partial_t \boldsymbol{g}|_{\boldsymbol{g}^*} = 0$. For the fermionic systems in $2<d<4$ dimensions considered here, they separate the theory space into a domain of attraction of the Gaussian fixed point $\boldsymbol{g}^* = 0$, and the strong-coupling regime, where the flow becomes unstable, signaled by a divergence of the renormalized couplings at finite RG scale. 
In the vicinity of a fixed point, the RG flow can be linearized and is then governed by the stability matrix $B_{ij} \equiv \partial (\partial_t g_i) / \partial g_j |_{\boldsymbol{g}^*}$. Let $\boldsymbol{v}^I$ be the eigenvectors and $-\Theta^I$ the eigenvalues of the stability matrix, i.e., $B \boldsymbol{v}^I = - \Theta^I \boldsymbol{v}^I$ with $I = 1,\dots, 4$ ($I = 1,2, 3$) for $\Nf \geq 2$ ($\Nf=1$). In the basis $\{ \boldsymbol{v}^I\}$ we can thus integrate the linearized flow
\begin{align}
\boldsymbol{v}^I(t) =  \boldsymbol{v}^I(0) \exp(-\Theta^I t),
\end{align}
and $\boldsymbol{v}^I$ therewith defines a RG relevant (irrelevant) direction, if the exponent $\Theta^I > 0$ ($\Theta^I < 0$). We call a fixed point ``critical'', if exactly one exponent is positive and all others are negative. Such critical fixed points may be associated to second-order phase transitions.

The quadratic form of the one-loop beta functions [Eq.~\eqref{eq:beta-general}] causes the flow to exhibit the following simplifying properties:

\begin{enumerate}[(1)]
\item Any nontrivial fixed point $\boldsymbol{g}^* \neq 0$ has a positive critical exponent $\Theta = d-2$. The corresponding RG relevant direction is given by the fixed-point vector $\boldsymbol{g}^*$ itself. 
\item The (straight) line connecting the Gaussian with a non-Gaussian
fixed point is invariant under the RG.
\item A plane that contains the Gaussian fixed point $\boldsymbol{g}^*_\mathcal O = 0$ and three non-Gaussian fixed points $\boldsymbol{g}_{\mathcal A}^*, \boldsymbol{g}_{\mathcal B}^*, \boldsymbol{g}_{\mathcal C}^* \neq 0$ is RG-invariant if $\boldsymbol{g}_{\mathcal A,\mathcal B,\mathcal C}^*$ are pairwise linear independent. Similar relations hold for higher-dimensional subspaces.
\end{enumerate}
These properties are readily be shown by making use of
Eq.~\eqref{eq:beta-general}. The stability matrix has the form $B_{ij}
= (d-2) \delta_{ij} + 2 \sum_k g_k^* A_{i,kj}$ and thus \cite{Gies:2003dp}
\begin{align}
\sum_j B_{ij} g_j^* = (d-2) g_i^* + 2 \sum_{k,j} g_k^* A_{i,kj} g_j^* = -(d-2)
g_i^*,
\end{align}
where in the last step we have made use of the fixed-point equation $\partial_t \boldsymbol{g}|_{\boldsymbol{g}^*} = 0$. This proves~(1). Property~(2) is shown similarly: Let $\lambda \boldsymbol{g}^*$ with $\lambda \in \mathbbm{R}$ parametrize the line connecting $\mathcal O$ with the non-Gaussian fixed point at $\boldsymbol{g}^* \neq 0$. Then 
\begin{align}
\partial_t (\lambda \boldsymbol g^*) = (d-2) (1 - \lambda) (\lambda \boldsymbol g^*)
\end{align}
and is thus parallel to $\boldsymbol g^*$ itself. Ad (3): Without loss of generality we can define a basis with the two fixed-point vectors $\boldsymbol{g}_{\mathcal A,\mathcal B}^*$ being the first two basis elements and all other basis vectors pointing out of this plane. The flow out of the plane is thus
\begin{align} \label{eq:proof-2}
\partial_t g_i \Bigr|_{g_{m\geq 3} = 0} = (d-2) g_i + \sum_{j,l\leq 2} g_j
A_{i,jl} g_l, \qquad i \geq 3.
\end{align}
Plugging the three (linear independent) fixed-point conditions $\partial_t
\boldsymbol{g}_{\mathcal A,\mathcal B,\mathcal C}^*$ into Eq.~\eqref{eq:proof-2}
gives $A_{i,11} = A_{i,12} = A_{i,22} = 0$. The plane containing $\mathcal O,
\mathcal A, \mathcal B$, and $\mathcal C$ is thus RG invariant. This reasoning
can be generalized for $n$-dimensional subspaces containing $n(n+1)/2$
non-Gaussian (and suitably located) fixed points.

\section{Fixed-point structure for \texorpdfstring{$\Nf = 1$}{Nf=1}}
\label{sec:fixed-points-Nf-1}

\begin{figure}[t]
\includegraphics[width=.48\textwidth]{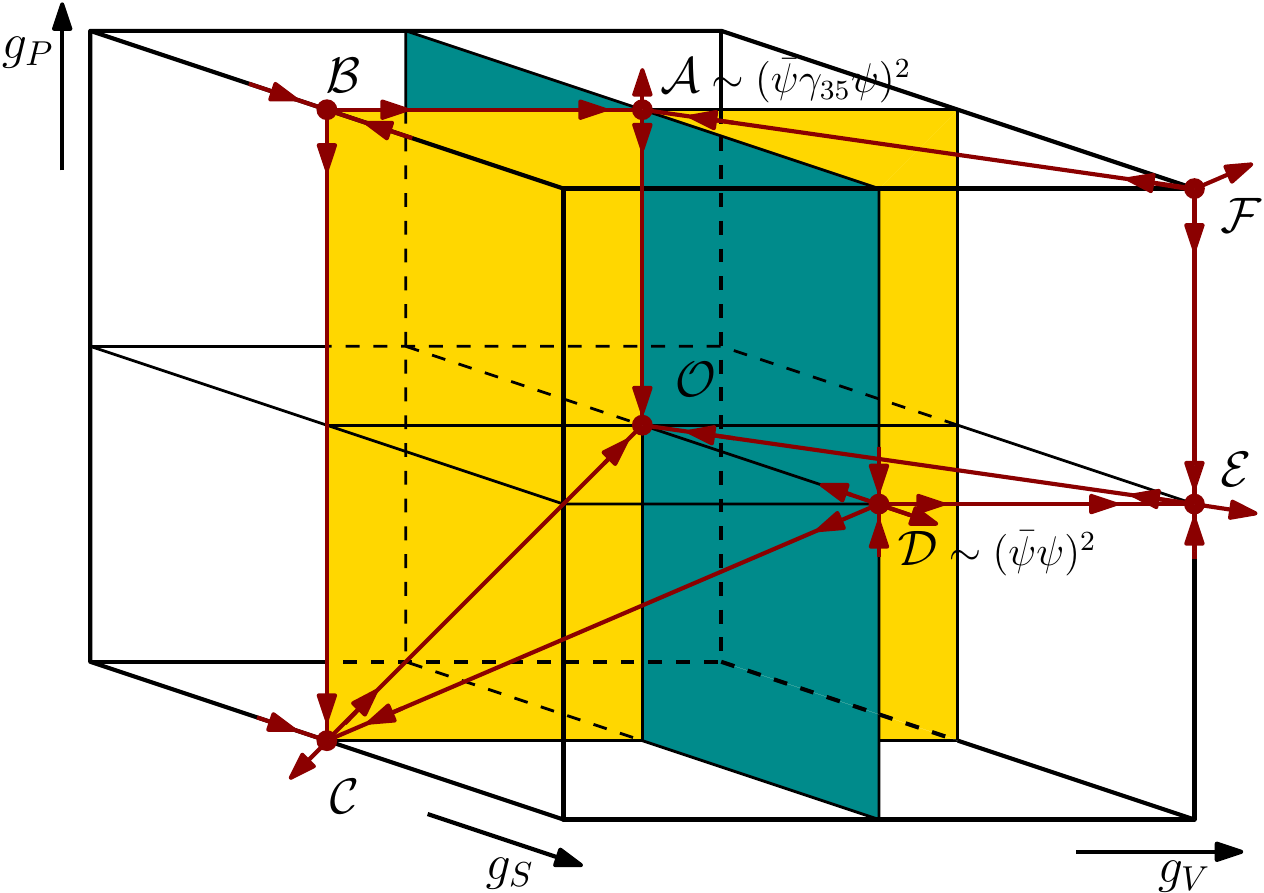}
\caption{Schematic fixed-point structure for $\Nf=1$ (not true to scale!). Any non-Gaussian fixed point $\boldsymbol g^* \neq 0$ has a relevant direction along $\boldsymbol g^*$ itself. There are three critical fixed points $\mathcal A$, $\mathcal C$, and $\mathcal E$ with exactly one relevant direction. The fixed points $\mathcal A$, $\mathcal B$, and $\mathcal C$ are located in the higher-symmetric $\mathrm U(2\Nf)$ space $g_S = 0$, which is invariant under the RG (yellow/light gray).  $\mathcal A$ and $\mathcal D$ [together with the fixed point $\mathcal{G}$ at $(g_S,g_V,g_P) = (\infty,0,\infty)$] are located in the invariant plane $g_V=0$ (cyan/dark gray). }
\label{fig:fixed-points-Nf1}
\end{figure}

For the sake of clarity, we perform the following fixed-point analysis
in $d=3$ space-time dimensions. Note, however, that within our
approximation we obtain the very same results in general $2<d<4$ upon
the appropriate rescaling of fixed-point values $g \mapsto (d-2) g$
and critical exponents $\Theta \mapsto (d-2) \Theta$.  We start with
the one-flavor case $\Nf=1$, for which the additional ``Fierz''
identity renders the theory space three-dimensional. Out of the
$2^3=8$ possibly degenerate and complex solutions of the fixed-point
equations [Eqs.~\eqref{eq:flow-Nf1-A}--\eqref{eq:flow-Nf1-C}] we find
$7$ real and distinct fixed points $\mathcal O$, $\mathcal A$,
$\mathcal B$, $\mathcal C$, $\mathcal D$, $\mathcal E$, and $\mathcal
F$.  Their locations are depicted schematically in
Fig.~\ref{fig:fixed-points-Nf1}.  From the equations for general $\Nf
>1$ [Eqs.~\eqref{eq:flow-A}--\eqref{eq:flow-D}] we infer that the
$8$th solution $\mathcal {G}$ of the fixed-point equations, present
for $\Nf > 1$, diverges for $\Nf \to 1$. Fixed points $\mathcal O$,
$\mathcal A$, $\mathcal B$, and $\mathcal C$ lie in the
two-dimensional, and thus RG-invariant, subspace with $g_S = 0$
(yellow/light gray plane in Fig.~\ref{fig:fixed-points-Nf1}).  This
plane is in fact the Thirring subspace with $\mathrm U(2\Nf)$
symmetry. The fixed-point structure of the $\mathrm U(2\Nf)$ theory
space has been investigated earlier~\cite{GiesJanssen2010,
  Mesterhazy2012, JanssenGies2012}, showing that $\mathcal A$ and
$\mathcal C$ are critical fixed points with exactly one relevant
direction \textit{within this plane}. $\mathcal A$ is the fixed point
which describes the Gross-Neveu model in the irreducible
two-dimensional representation of the Clifford algebra with a
four-fermi interaction that can be rewritten in terms of two-component
(Weyl) spinors $\chi$ as $(P)^2 = (\bar\psi\gamma_{35}\psi)^2 \propto
(\bar\chi\chi)^2$ [Eq.~\eqref{eq:P2-barchi-chi}]. The critical
behavior of $\mathcal A$ is well-known \cite{RosensteinWarrPark1989,
  RosensteinYuKovner1993, Gracey1991,
  VasilievDerkachovKivelStepanenko1993, HandsKocicKogut1993,
  RosaVitaleWetterich2001}. The fixed point $\mathcal C$ describes the
chiral phase transition expected in the three-dimensional Thirring
model and as such has been dubbed Thirring fixed point
\cite{GiesJanssen2010, JanssenGies2012, Mesterhazy2012}---although we
emphasize that $\mathcal C$ does not lie on the Thirring axis with
pure $(V)^2=(\bar\psi\gamma_\mu\psi)^2$ interaction, but its
fixed-point action includes both $(P)^2$ and $(V)^2$
contributions~\cite{GiesJanssen2010}. The critical behavior of
$\mathcal C$ has recently also been approached~\cite{JanssenGies2012,
  Mesterhazy2012}. Here, we are able to determine whether or not small
perturbations \textit{out} of the $\mathrm U(2\Nf)$-symmetric plane
are relevant in the sense of the RG. From
Eqs.~\eqref{eq:flow-Nf1-A}--\eqref{eq:flow-Nf1-C} we find the
following critical exponents:
\begin{align}
\mathcal A: \quad \Theta & = \left( 1, -2, -2  \right), \\
\mathcal C: \quad \Theta & = \left( 1, -3 + \sqrt{5}, - 3 \sqrt{5} + 5 \right),
\end{align}
and therefore both $\mathrm U(2\Nf)$ critical fixed points remain
critical also in the lower-symmetric $\mathrm U(\Nf) \times \mathrm
U(\Nf)$ theory space. In the vicinity of both $\mathcal A$ at $(g_S^*,
g_V^*, g_P^*) = (0, 0, \frac{1}{2})$ and $\mathcal C$ at $(0, -
\frac{3-\sqrt{5}}{4}, - \frac{\sqrt{5}-1}{4})$ any $\mathrm
U(2\Nf)$-breaking but $\mathrm U(\Nf) \times \mathrm
U(\Nf)$-preserving perturbation is thus RG irrelevant for $\Nf=1$,
since the only relevant directions of each fixed point are the
fixed-point vectors $\boldsymbol{g}^*_\mathcal{A}$ and
$\boldsymbol{g}_\mathcal{C}^*$ itself.

\begin{table}[t]
\caption{Fixed points for $\Nf=1$, their locations, and number of relevant directions. GN: Gross-Neveu.}
\label{tab:fixed-points-Nf1}
\begin{tabular*}{.48\textwidth}{@{\extracolsep{\fill} }cccc}
\hline\hline
& $(g_S^*, g_V^*,g_P^*)$ & $\#(\Theta>0)$ \\
\hline
$\mathcal O$ & $(0,0,0)$ & 0 & Gaussian\\
$\mathcal A$ & $(0,0,1/2)$ & 1 & irreducible GN\\
$\mathcal B$ & $(0, - \frac{3+\sqrt{5}}{4}, \frac{\sqrt{5}+1}{4})$ & 2\\
$\mathcal C$ & $(0, - \frac{3-\sqrt{5}}{4}, - \frac{\sqrt{5}-1}{4})$ & 1 & Thirring\\
$\mathcal D$ & $(1/2,0,0)$ & 2 & reducible GN\\
$\mathcal E$ & $(\frac{3(3 - \sqrt{5})}{4} , \frac{3 - \sqrt{5}}{4}, 
 \frac{5 - 2 \sqrt{5}}{2})$ & 1\\
$\mathcal F$ & $(\frac{3(3 + \sqrt{5})}{4} , \frac{3 + \sqrt{5}}{4}, 
 \frac{5 + 2 \sqrt{5}}{2})$ & 2\\
$\mathcal {G}$ & $(\infty, 0, \infty)$ & $3$ & diverges for $\Nf\to 1$\\
\hline\hline
\end{tabular*}
\end{table}

A second RG-invariant plane is given by $g_V = 0$ (cyan/dark gray
plane in Fig.~\ref{fig:fixed-points-Nf1}), and it contains besides
$\mathcal A$ the non-Gaussian fixed point $\mathcal D$ at $(g_S^*,
g_V^*, g_P^*) = (\frac{1}{2}, 0, 0)$. $\mathcal D$ describes a theory
with pure $(\bar\psi\psi)^2$ four-fermi interaction, i.e., the
Gross-Neveu model in the reducible, four-dimensional, representation
of the Clifford-Algebra \cite{BraunGiesScherer2011}. The $8$th
solution $\mathcal {G}$ of the fixed-point equations that diverges for
$\Nf \to 1$ also lies in this subplane at $(g,0,g)$ with $g \propto
(\Nf-1)^{-1}$. The line $(g_S, g_V, g_P) = \lambda (1,0,1)$ is
therefore also RG invariant, just as the Gross-Neveu axes $\lambda
(1,0,0)$ and $\lambda (0,0,1)$. As has been observed previously
\cite{HerbutJuricicRoy2009}, both $\mathcal A$ and $\mathcal D$ have
only one relevant direction \textit{in this plane}. However, while
perturbations out of this plane are irrelevant in the vicinity of
$\mathcal A$, such is not the case in the vicinity of fixed point
$\mathcal D$: For this fixed point the critical exponents are
\begin{align}
\mathcal D: \quad \Theta & = \left(1, 2/3, -2 \right),
\end{align}
and thus the reducible-Gross-Neveu fixed point $\mathcal D$ has \textit{two}
relevant directions in the $\mathrm U(\Nf) \times \mathrm U(\Nf)$ theory space
for $\Nf=1$. Since there is no higher symmetry that could forbid perturbations
$\propto g_V$ the fixed point $\mathcal D$ is not a critical fixed point for
$\Nf = 1$ and cannot describe a second-order phase transition, as long as only
one microscopic parameter is tuned.

There is, however, a third critical fixed point, located at
\begin{align} \label{eq:fixed-point-E}
\mathcal E: \quad (g_S^*,g_V^*,g_P^*) = \left( \frac{3(3 - \sqrt{5})}{4} , \frac{3 - \sqrt{5}}{4}, 
 \frac{5 - 2 \sqrt{5}}{2} \right),
\end{align}
which has the same critical exponents as fixed point $\mathcal C$ in our approximation,
\begin{align}
\mathcal E: \quad \Theta & = \left( 1, -3 + \sqrt{5}, - 3 \sqrt{5} + 5\right).
\end{align}
Small perturbations near fixed point $\mathcal D$ drive the flow to 
either $\mathcal C$ or $\mathcal E$, depending on the sign of $g_V$.
We have summarized our results for the $\Nf=1$ fixed-point structure in
Tab.~\ref{tab:fixed-points-Nf1}.

\section{Fixed-point structure for \texorpdfstring{$\Nf \geq 2$}{Nf >= 2}}
\label{sec:fixed-points-Nf-general}
\subsection{Collision of fixed points}
For $\Nf \geq 2$ the additional flow equation renders the $\mathrm U(\Nf) \times 
\mathrm U(\Nf)$ theory space four-dimensional, thus generating a more complex 
fixed-point structure. Generically, one expects $2^4=16$ possibly complex or 
degenerate solutions of the fixed-point equations 
[Eqs.~\eqref{eq:flow-A}--\eqref{eq:flow-D}]. We find that the number of real 
(and therefore physical) fixed points actually depends on the flavor number 
$\Nf$. In the small-$\Nf$ regime $2\leq \Nf < \Nf^{(1)}$ with
\begin{align*}
\Nf^{(1)} & = 
-\tfrac{4}{9}+\tfrac{1}{9}\left(9872-144\sqrt{3345}\right)^{\frac{1}{3}}
\\ & \quad 
+\tfrac{2}{9}\left(1234+18\sqrt{3345}\right)^{\frac{1}{3}} \\ 
& \simeq 3.76
\end{align*}
there are $12$ distinct real fixed points and two pairs of complex
conjugate solutions. Above $\Nf \geq \Nf^{(1)}$ both pairs
simultaneously become real and we find $16$ real fixed points. 
A selection of them is listed in Tab.~\ref{tab:fixed-points-Nf2}.
For particular values of $\Nf$ these solutions become degenerate, i.e.,
fixed points approach each other in coupling space as a function of
$\Nf$ and eventually collide. In general, when two fixed points
collide, two principally different situations are possible: (1) The
fixed points can ``run through'' each other as a function of $\Nf$ and
exchange roles with respect to the RG stability of the axis connecting
the two fixed points. This phenomenon is well known for the
Wilson-Fisher fixed point in $\phi^4$ theory as a function of
space-time dimension $d$, which collides with the Gaussian fixed point
for $d \nearrow 4$ and becomes unstable for $d>4$
\cite{Herbut2007}. It is also observed for the multicritical fermionic
fixed point on graphene's honeycomb lattice as a function of flavor
number $\Nf$~\cite{ClassenJanssenHerbutScherer2015}. (2) The other
possible situation is that the fixed points ``merge'' and eventually
disappear into the complex plane. Such has been observed in
many-flavor QCD \cite{GiesJaeckel2006}, three-dimensional
scalar~\cite{HalperinLubenskyMa1974} or
fermionic~\cite{KavehHerbut2005} QED, and recently also in
nonrelativistic systems with quadratic band touching
\cite{HerbutJanssen2014}.

\begin{table}[t]
\caption{Selected fixed points for $\Nf \geq 2$, their locations, and number of relevant directions. At $\Nf \searrow \Nf^{(1)}{} \simeq 3.76$ the three fixed points $\mathcal H$, $\mathcal I$, and $\mathcal J$ collide. $\mathcal I$ and $\mathcal J$ merge and subsequently become complex for $\Nf < \Nf^{(1)}$, and exchange stability of one direction with the fixed point $\mathcal H$. At $\Nf \nearrow \Nf^{(2)} = 6$, $\mathcal I$ collides with the Thirring fixed point $\mathcal C$, and $\mathcal J$ collides with $\mathcal K$, again exchanging roles: for $\Nf > 6$, $\mathcal C$ and $\mathcal K$ become critical fixed points. $\mathcal A$ and $\mathcal D$ on the two (irreducible and reducible) Gross-Neveu (GN) axes $\sim (\bar\psi\gamma_{35}\psi)^2$ and $(\bar\psi\psi)^2$, respectively, are critical fixed points for all $\Nf \geq 2$.}
\label{tab:fixed-points-Nf2}
\begin{tabular*}{.48\textwidth}{@{\extracolsep{\fill} }cccc}
\hline\hline
& $(g_S^*, g_P^*,g_V^*, g_T^*)$ & $\#(\Theta>0)$ \\
\hline
$\mathcal A$ & $(0,a,0,0)$ & $1$ & irreducible GN\\
$\mathcal B$ & $(0,b_1,b_2,0)$ & 
$\begin{cases} 
 3, & \Nf < 6 \\
 2, & \Nf > 6
\end{cases}$ \\
$\mathcal C$ & $(0,c_1,c_2,0)$ & 
$\begin{cases} 
 2, & \Nf < 6 \\
 1, & \Nf > 6
\end{cases}$
& Thirring\\
$\mathcal D$ & $(a,0,0,0)$ & $1$ & reducible GN \\
$\mathcal G$ & $(g, g, 0, 0)$ & 2 \\
$\mathcal H$ & $(h_1, h_1, h_2, h_2)$ &
$\begin{cases}
 1, & \Nf < \Nf^{(1)} \\
 2, & \Nf > \Nf^{(1)}
\end{cases}$ \\
$\mathcal I$ & $(i_1,i_2,i_3,i_4)$ &
$\begin{cases}
 1, & \Nf^{(1)} < \Nf < 6 \\
 2, & \Nf > 6
\end{cases}$ \\
$\mathcal J$ & $(i_1, i_2, i_4, i_3)$ &
$\begin{cases}
 1, & \Nf^{(1)} < \Nf < 6 \\
 2, & \Nf > 6
\end{cases}$ \\
$\mathcal K$ & $(0,c_1,0,c_2)$ & 
$\begin{cases} 
 2, & \Nf < 6 \\
 1, & \Nf > 6
\end{cases}$ \\
$\mathcal L$ & $(l_1, l_1, l_2, l_2)$ & $4$ \\
\hline\hline
\end{tabular*}
\end{table}

In our system, we find two values of $\Nf$ for which a collision of
fixed points occurs. The first one is particularly interesting from a
general viewpoint: In the limit $\Nf \searrow \Nf^{(1)}$ we in fact
find two \textit{triples} of solution, each consisting of
\textit{three} non-Gaussian fixed points that approach each other in
coupling space and eventually collide simultaneously at two different
locations. Let us discuss one of these triples in more detail. It
consists of the three fixed points $\mathcal H$, $\mathcal I$, and
$\mathcal J$, which for general $\Nf > \Nf^{(1)}$ are located at
\begin{align}
 \mathcal H: \quad (g_S^*,g_P^*,g_V^*, g_T^*) & = \left(h_1, h_1, h_2, h_2 \right), \\
 \mathcal I: \quad (g_S^*,g_P^*,g_V^*, g_T^*) & = \left(i_1, i_2, i_3, i_4 \right), \\
 \mathcal J: \quad (g_S^*,g_P^*,g_V^*, g_T^*) & = \left(i_1, i_2, i_4, i_3 \right), 
\end{align}
where $h_{1,2} \equiv h_{1,2}(\Nf)$, $i_{1,\dots,4} \equiv
i_{1,\dots,4}(\Nf)$ ($|i_3|\geq|i_4|$) are functions of $\Nf$, which
we do not display explicitly for reasons of readability. The symmetry
between the locations of $\mathcal I$ and $\mathcal J$ is determined
by the $g_V \leftrightarrow g_T$ symmetry of the one-loop flow
equations. For $\Nf \searrow \Nf^{(1)}$ we find $i_{1,2} \to h_1$ and
$i_{3,4} \to h_2$, i.e., all three fixed points collide. Two of them
($\mathcal I$ and $\mathcal J$) merge and disappear into the complex
plane for $\Nf < \Nf^{(1)}$, while the third ($\mathcal H$) remains as
a real fixed point. As can be read off from
Tab.~\ref{tab:fixed-points-Nf2}, $\mathcal I$ and $\mathcal J$ are
critical fixed points with exactly one RG relevant direction for $\Nf$
above (but close to) $\Nf^{(1)}$ and exchange one stable direction
with the third fixed point $\mathcal H$, which becomes critical only
for $\Nf < \Nf^{(1)}$. Thus, at this \textit{triple collision} both
phenomena, merging and disappearing into complex plane as well as
exchanging roles with respect to RG stability, are realized. The
analogous behavior can be simultaneously observed for the second
triple of fixed points with $2$ respectively $3$ relevant directions
(these fixed points are not among those explicitly detailed in
Tab.~\ref{tab:fixed-points-Nf2}).  To our knowledge, the present
$\mathrm U(\Nf) \times \mathrm U(\Nf)$ system represents the
first-known example displaying such triple collision.

The second value of $\Nf$ at which we find degenerate fixed-point solutions is
\begin{align} \label{eq:Nfcrit2}
 \Nf^{(2)} = 6.
\end{align}
Here, two simultaneous collisions of the first kind (\`a la
Wilson-Fisher in $d\nearrow 4$) occur: For $\Nf \nearrow \Nf^{(2)}$ we
find that fixed point $\mathcal I$ collides with the Thirring fixed
point
\begin{align}
 \mathcal C: \quad (g_S^*,g_P^*,g_V^*, g_T^*) = \left(0, c_1, c_2, 0 \right),
\end{align}
with the coordinates $c_{1,2} \equiv c_{1,2}(\Nf) < 0$, i.e., $i_1 = 0$, $i_2=c_1$, $i_3 = c_2$, and $i_4 = 0$ for $\Nf=\Nf^{(2)}$. The Thirring fixed point, which lies in the higher-symmetric $\mathrm U(2\Nf)$ subspace, has two relevant directions for $2\leq \Nf < 6$ and exchanges roles with respect to stability of one of its RG relevant directions with fixed point $\mathcal I$ for $\Nf = \Nf^{(2)} = 6$. Only for $\Nf > \Nf^{(2)}$ (and $\Nf = 1$, see above) the Thirring fixed point is thus a critical fixed point, with $\mathcal I$ being critical below (but close to) $\Nf^{(2)}$ and developing a second relevant direction above $\Nf^{(2)}$. The same behavior can be observed for the fixed point $\mathcal J$ that collides and exchanges its role with respect to RG stability with the fixed point
\begin{align}
 \mathcal K: \quad (g_S^*,g_P^*,g_V^*, g_T^*) = \left(0, c_1, 0, c_2 \right),
\end{align}
simultaneously at $\Nf = \Nf^{(2)}$. We note that the simultaneous
collision of \textit{two} pairs of fixed points, just as the fact that
both triples collide at the same $\Nf^{(1)}$ as discussed above, is a
consequence of the invariance of the flow equations
[Eqs.~\eqref{eq:flow-A}--\eqref{eq:flow-D}] under the exchange of
$g_V$ and $g_T$, which ensures that if $(g_S^*,g_P^*,g_V^*, g_T^*)$
solves the fixed-point equations, so does $(g_S^*,g_P^*,g_T^*,
g_V^*)$, and both (not necessarily distinct) fixed points have the
same critical exponents.  We discuss the stability of the $\mathrm
U(2\Nf)$ subspace in the vicinity of the Thirring fixed point in
detail in subsection~\ref{sec:stability-thirring}.

Just as in the one-flavor case, the two (irreducible and reducible) Gross-Neveu axes given by pure $(\bar\psi\gamma_{35}\psi)^2$ and $(\bar\psi\psi)^2$ interactions, respectively, each define a RG invariant one-dimensional subspace, since they contain the non-Gaussian fixed points
\begin{align}
 \mathcal A: \quad (g_S^*,g_P^*,g_V^*, g_T^*) = \left(0, a, 0, 0 \right)
\end{align}
and
\begin{align}
 \mathcal D: \quad (g_S^*,g_P^*,g_V^*, g_T^*) = \left(a, 0, 0, 0 \right),
\end{align}
respectively, where $a \equiv a(\Nf) > 0$. Now, for $\Nf \geq 2$, both $\mathcal A$ and $\mathcal D$ are critical fixed points with exactly one relevant direction---in contrast to the one-flavor case, where $\mathcal D$ exhibited two relevant directions. All other fixed points, not listed in Tab.~\ref{tab:fixed-points-Nf2}, have two or more relevant directions for all $\Nf \geq 2$.

Let us briefly summarize the results we have obtained so far (cf.\ 
Tab.~\ref{tab:fixed-points-Nf2}): For $2 \leq \Nf < \Nf^{(1)} \simeq 3.76$ there 
are three critical fixed points $\mathcal A$, $\mathcal D$, and $\mathcal H$. A 
pair $\mathcal I$ and $\mathcal J$ emerges from the complex plane at the 
location of $\mathcal H$ when $\Nf = \Nf^{(1)}$. For $\Nf^{(1)}<\Nf<\Nf^{(2)}$ 
there are then four critical points $\mathcal A$, $\mathcal D$, $\mathcal I$, 
and $\mathcal J$. At $\Nf^{(2)}=6$ the fixed points $\mathcal I$ and $\mathcal 
J$ collide with $\mathcal C$ and $\mathcal K$, respectively. For $\Nf > 
\Nf^{(2)}$ we still have four critical fixed points, which however are now 
$\mathcal A$, $\mathcal D$, $\mathcal C$, and $\mathcal K$.

\subsection{RG invariant subspaces}
\begin{figure*}
 \includegraphics[width=.32\textwidth]{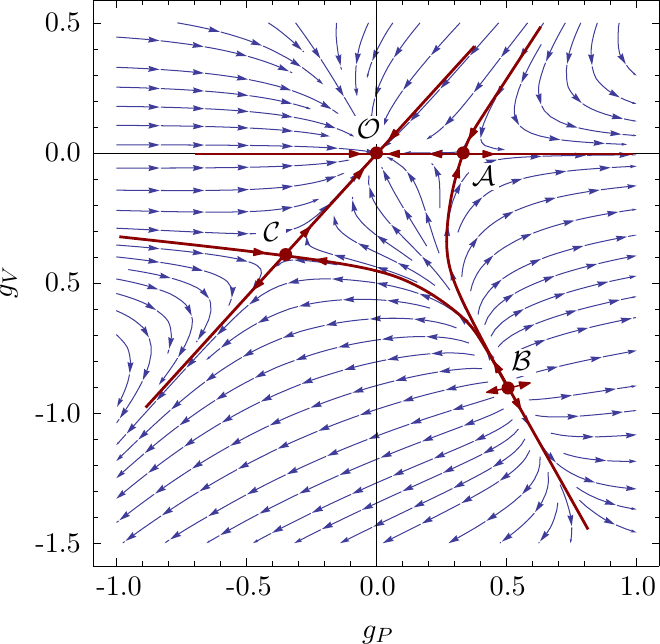}\hfill
 \includegraphics[width=.32\textwidth]{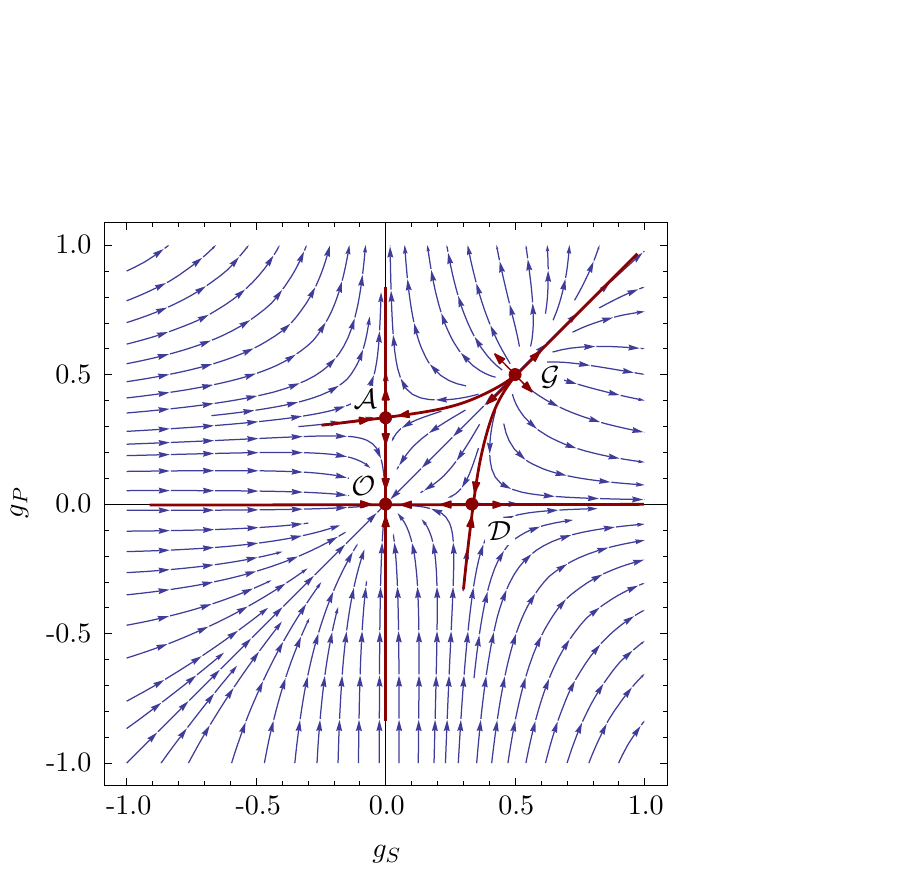}\hfill
 \includegraphics[width=.32\textwidth]{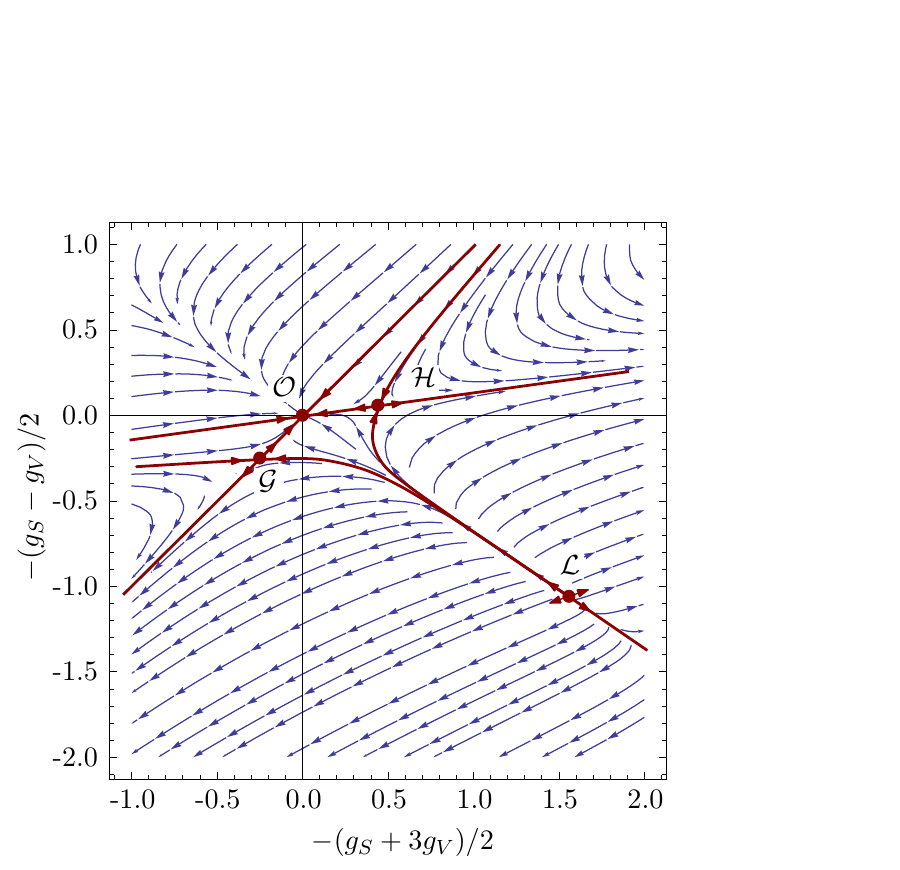}
 \caption{Flow in RG invariant planes for $\Nf = 2$. Arrows point
   towards the IR. Left: Thirring subspace with higher $\mathrm
   U(2\Nf)$ symmetry for $g_S = g_T = 0$
   \cite{GiesJanssen2010}. Middle: Gross-Neveu subspace for $g_V = g_T
   = 0$. Right: subspace defined by the Fierz-transformed interactions
   $(S^D)^2$ and $(V^D)^2$, given by $g_S = g_P$ and $g_V = g_T$ and
   parametrized by the couplings $g_S^D \equiv - \frac{1}{2} (g_S + 3
   g_V)$ (horizontal axis) and $g_V^D \equiv - \frac{1}{2}(g_S - g_V)$
   (vertical axis).}
 \label{fig:rg-flow-invariant-planes}
\end{figure*}
The fixed-point structure can further be elucidated by considering the
flow in subspaces that are closed under the action of the RG. Besides
the RG-closed one-dimensional lines connecting each non-Gaussian fixed
point with the Gaussian fixed point, we find three RG invariant
two-dimensional planes, each consisting of three non-Gaussian fixed
points and the Gaussian fixed point. These are
\begin{enumerate}[(1)]
 \item the Thirring subspace, defined by $$(g_S, g_P, g_V, g_T) = (0, g_P, g_V, 0),$$ consisting of the fixed point $\mathcal A$, i.e., the non-Gaussian fixed point of the Gross-Neveu model in the irreducible representation of the Clifford algebra, $\mathcal B$ at $\boldsymbol{g}^* = (0, b_1, b_2, 0)$, and the Thirring fixed point $\mathcal C$,
 \item the Thirring subspace's equivalent for $g_V \leftrightarrow g_T$, given by $$(g_S, g_P, g_V, g_T) = (0, g_P, 0, g_T),$$
 consisting of fixed point $\mathcal A$, the Thirring fixed point's equivalent fixed point $\mathcal K$, and a not further specified fixed point at $\boldsymbol{g}^* = (0,b_1,0,b_2)$ (equivalent of $\mathcal B$),
 \item the Gross-Neveu subspace, with $$(g_S, g_P, g_V, g_T) = (g_S, g_P, 0, 0),$$ consisting of the fixed points $\mathcal A$ and $\mathcal D$ of the irreducible and reducible Gross-Neveu models, as well as the fixed point $\mathcal G$ at $g_S^* = g_P^*$,
 \item a not further specified subspace, in which $$(g_S, g_P, g_V, g_T) = (g_S, 
 g_S, g_V, g_V),$$ consisting of the fixed points $\mathcal H$, $\mathcal L$, 
 and $\mathcal G$. By means of the Fierz 
 identities~\eqref{eq:Fierz-A}--\eqref{eq:Fierz-D}, we can rewrite the 
 interactions in this subspace as a linear combination of the two 
 Fierz-transformed (``dual'') interactions $(S^D)^2$ and $(V^D)^2$:
 \begin{multline} \label{eq:fierz-sd-vd}
  g_S (S)^2 + g_S (P)^2 + g_V (V)^2 + g_V (T)^2 \\
  = - \frac{1}{2} (g_S + 3 g_V) (S^D)^2 - \frac{1}{2}(g_S - g_V) (V^D)^2.
 \end{multline}

\end{enumerate}
Subspaces (1) and (3) can be associated with the corresponding
one-flavor subspaces studied above. The RG flow in subspace (2) is
completely equivalent to the Thirring subspace (1). The invariant
$(S^D)^2$-$(V^D)^2$ subspace (4) is new. We have depicted the RG flow
within the invariant planes (1), (3), and (4) in
Fig.~\ref{fig:rg-flow-invariant-planes} for the case of $\Nf=2$.
While the flow in directions orthogonal to an invariant plane vanishes
\textit{per definitionem}, we emphasize that the answer to the
interesting question whether or not small perturbations out of the
plane are RG relevant depends on the location considered on the plane,
as well as on the flavor number $\Nf$. For instance, in the vicinity
of the fixed point $\mathcal H$ (cf.\ right panel of
Fig.~\ref{fig:rg-flow-invariant-planes}), which is a critical fixed
point for $2 \leq \Nf < \Nf^{(1)}$, small perturbations orthogonal to
the $(S^D)^2$-$(V^D)^2$-plane are RG irrelevant (relevant) for $2 \leq
\Nf < \Nf^{(1)}$ ($\Nf > \Nf^{(1)}$). By contrast, near the fixed
point $\mathcal L$, which for all $\Nf \geq 2$ has four relevant
directions, perturbations are always relevant. For all $\Nf \geq 2$
and in the vicinity of all four fixed points $\mathcal O$, $\mathcal
A$, $\mathcal D$, and $\mathcal G$, the Gross-Neveu plane $g_V = g_T =
0$ (middle panel of Fig.~\ref{fig:rg-flow-invariant-planes}) is stable
under small perturbations out of this plane. (Note that this was
\textit{not} the case near the fixed point $\mathcal D$ for $\Nf = 1$,
see above.) In the following we discuss in detail the stability of the
higher-symmetric $\mathrm U(2\Nf)$ subspace against symmetry-breaking
perturbations in the vicinity of the Thirring fixed point $\mathcal C$
(cf.\ left panel of Fig.~\ref{fig:rg-flow-invariant-planes}), as this
question has been under some debate in the past \cite{DelDebbioHandsMehegan1997, 
  ChristofiHandsStrouthos2007, JanssenGies2012, ChandrasekharanLi2012-b}.

\subsection{(In-)stability of Thirring subspace against symmetry-breaking perturbations}
\label{sec:stability-thirring}
\begin{figure*}
 \includegraphics[width=0.33\textwidth]{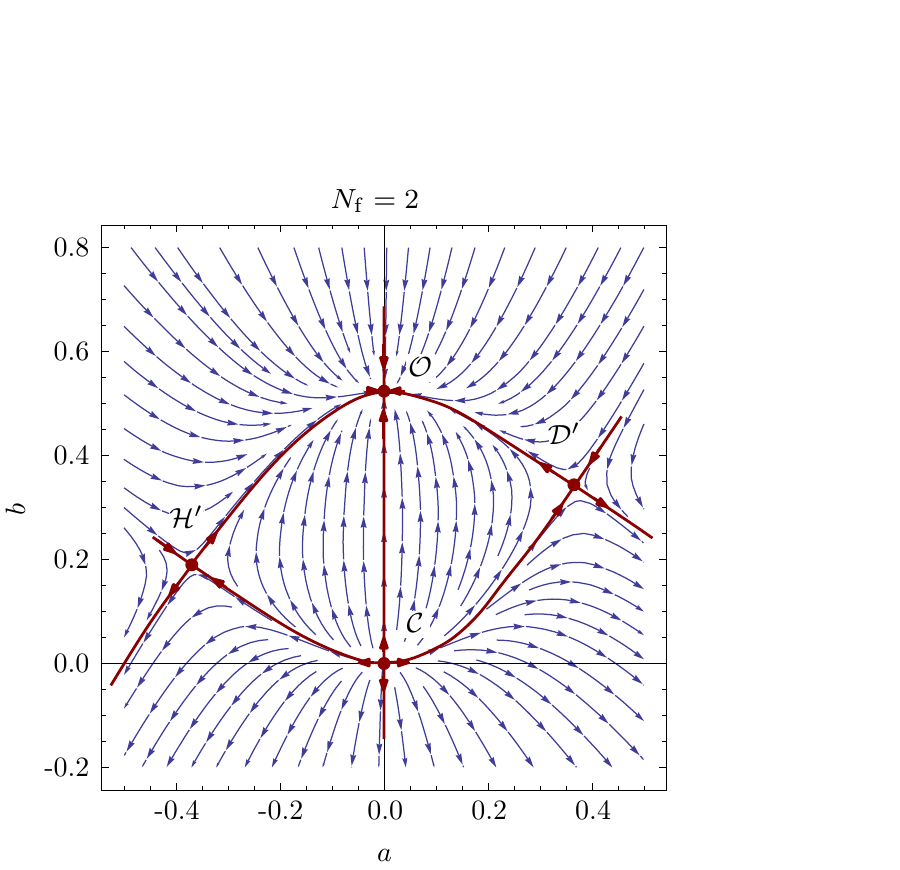}\hfill
 \includegraphics[width=0.33\textwidth]{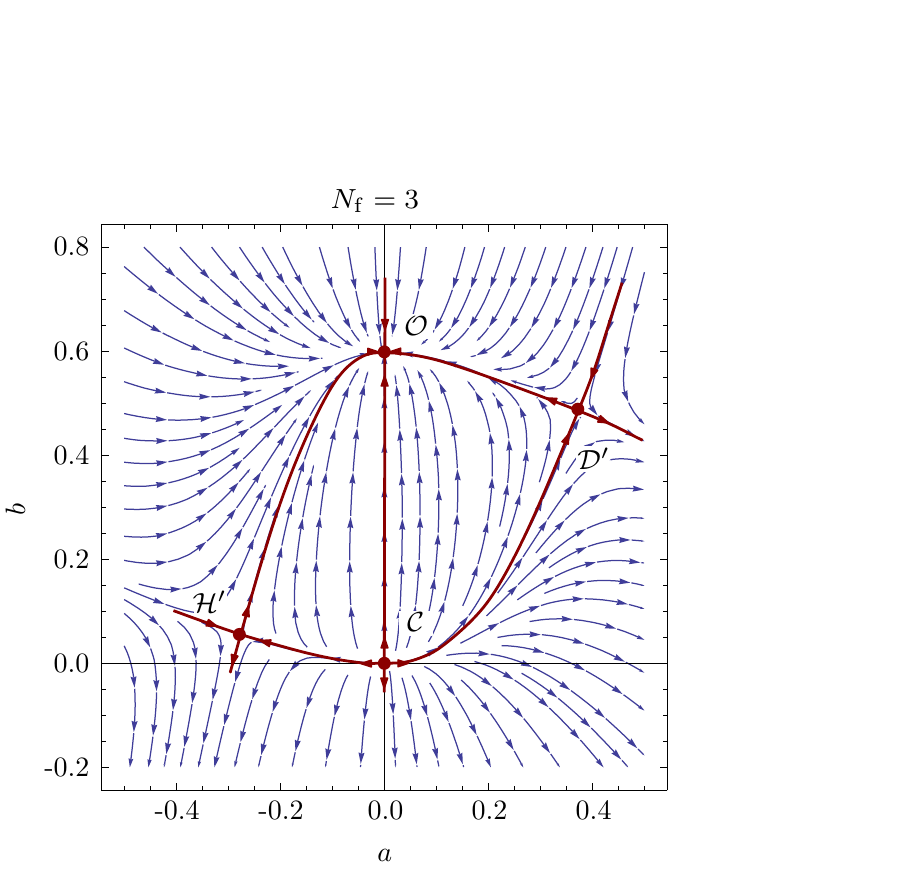}\hfill
 \includegraphics[width=0.33\textwidth]{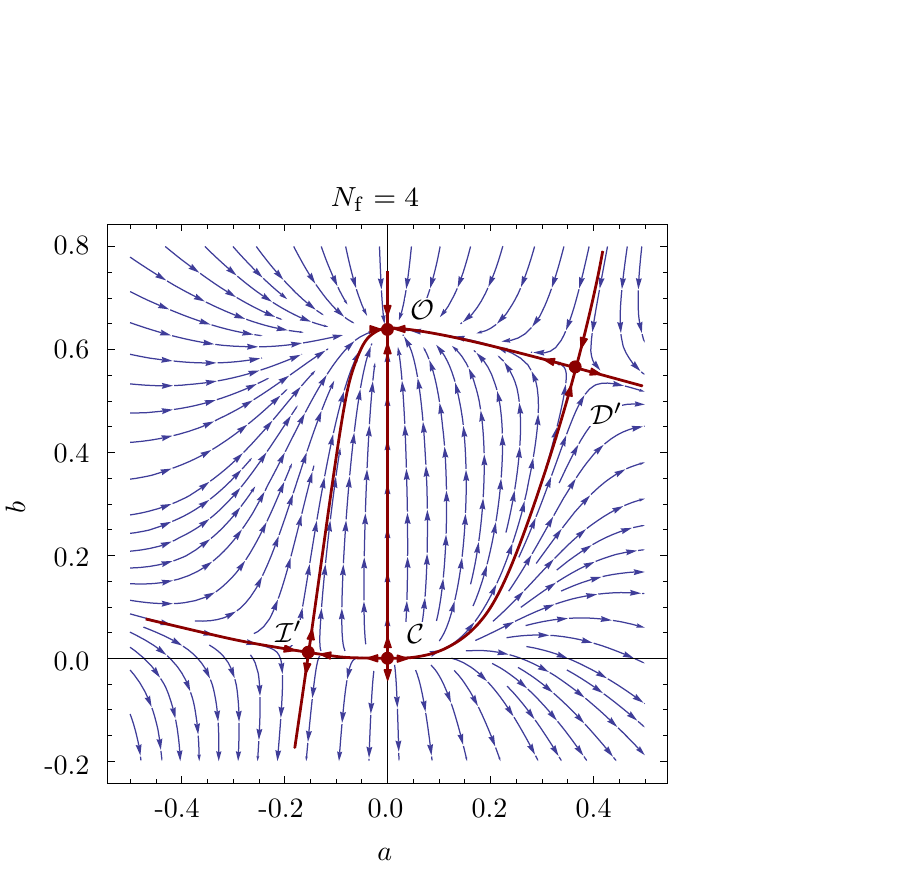}\\[1em]
 \includegraphics[width=0.33\textwidth]{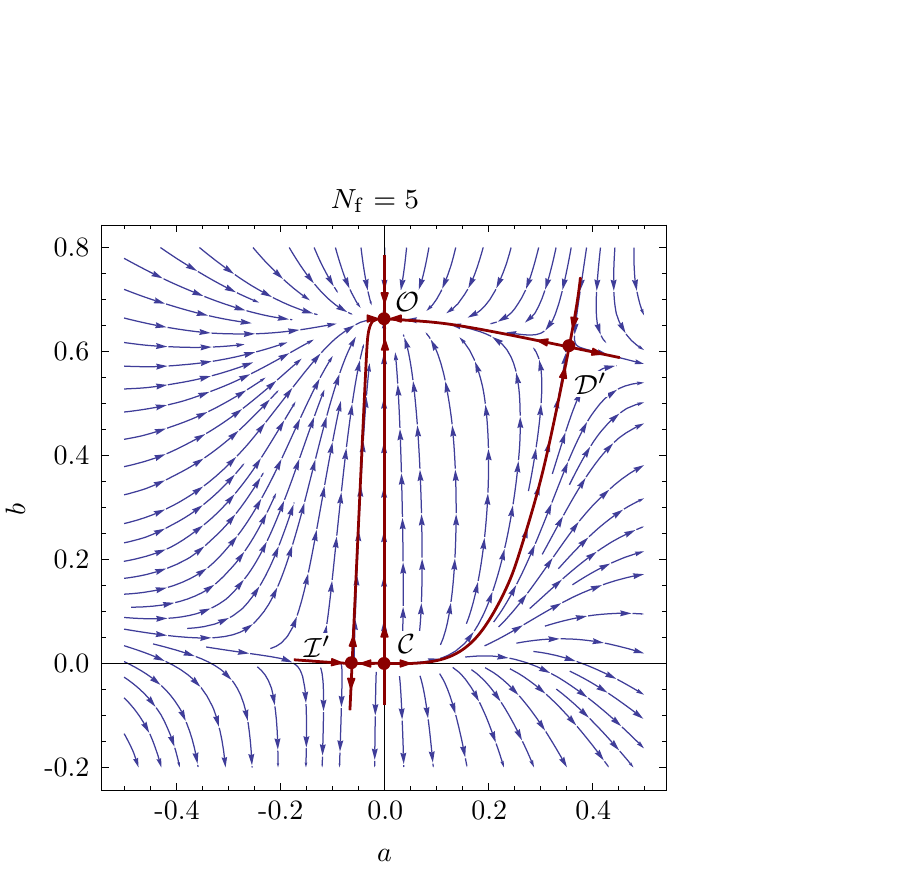}\hfill
 \hfill\includegraphics[width=0.33\textwidth]{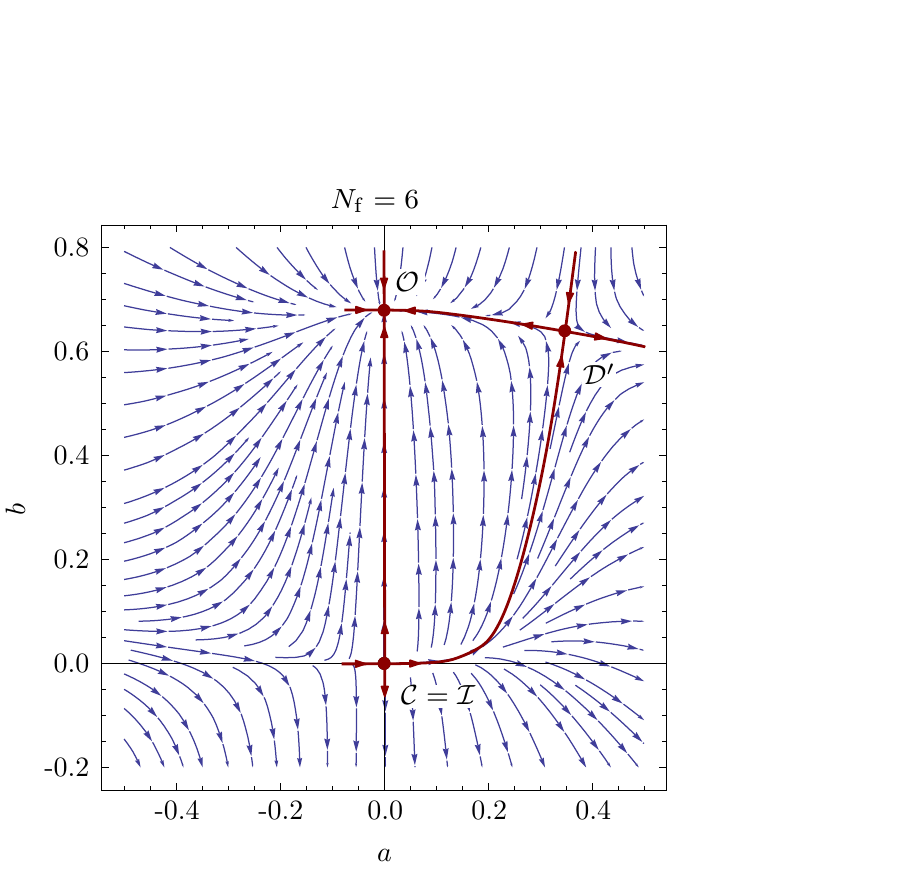}\hfill
 \includegraphics[width=0.33\textwidth]{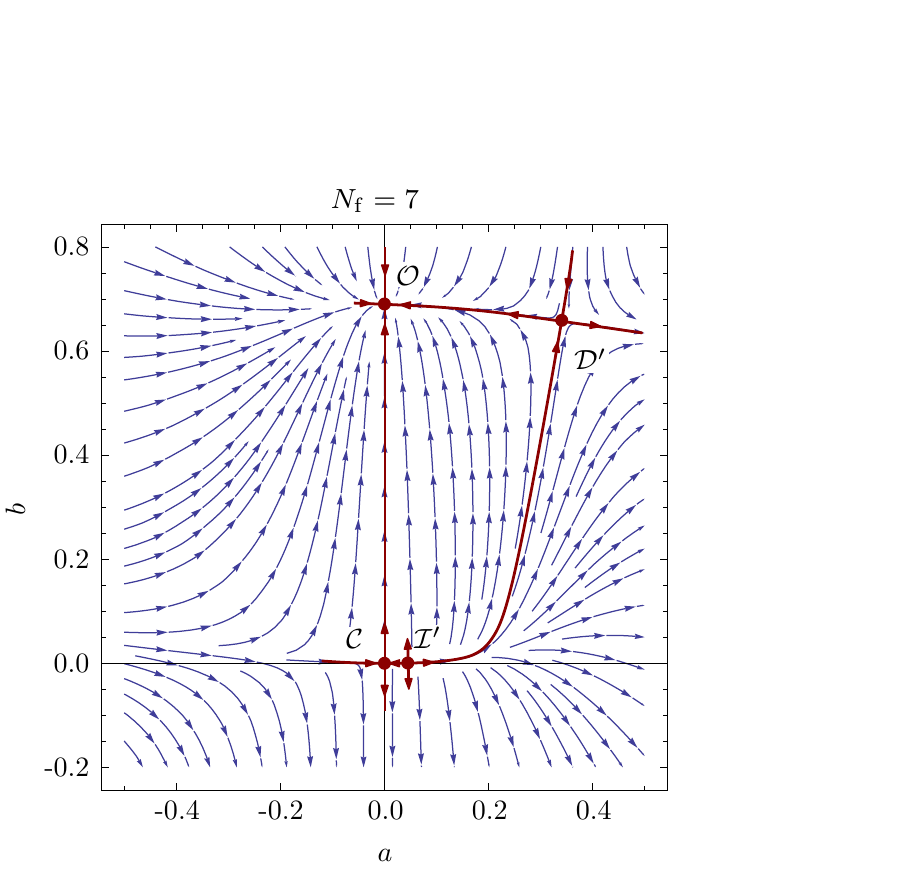}
 \caption{RG flow in the (RG noninvariant!) plane spanned by the
   Thirring fixed point $\mathcal C$ and the corresponding
   eigendirections $\boldsymbol{v_{1,2}}$ with the largest
   corresponding exponents $\Theta_{1,2}$. This plane is given by
   $\boldsymbol{g} = \boldsymbol{g}^*_\mathcal C + a \boldsymbol{v_1}
   + b \boldsymbol {v_2}$ with coordinates $a, b \in \mathbbm R$ and
   basis vectors $\boldsymbol{v_2} \propto \boldsymbol{g}^*_\mathcal
   C$ (corresponding exponent $\Theta_2 = 1$) and $\boldsymbol{v_1}$,
   the latter being the second relevant direction for $\Nf < 6$,
   becoming marginal for $\Nf = 6$, and eventually irrelevant for $\Nf
   > 6$. The vertical axis $a=0$ thus represents the Thirring
   subspace, while finite $a \neq 0$ corresponds to $\mathrm
   U(2\Nf)$-symmetry-breaking perturbations. By varying the starting
   points of the RG flow in the vicinity of $\mathcal C$ there is a
   certain finite region for which the flow runs into the Gaussian
   fixed point $\mathcal O$. Outside this region we find a runaway
   flow. At the boundaries the flow runs into a critical fixed point
   with exactly one relevant direction. This is the fixed point
   $\mathcal H$ ($\mathcal D$) if $a < 0$ ($a>0$) for $2\leq \Nf <
   \Nf^{(1)}$ (top left and middle), the fixed point $\mathcal I$
   ($\mathcal D$) if $a < 0$ ($a>0$) for $\Nf^{(1)} < \Nf < \Nf^{(2)}
   = 6$ (top right and bottom left), and eventually the Thirring fixed
   $\mathcal C$ itself (independent of $|a| \ll 1$) for $\Nf>
   \Nf^{(2)}$, which only then becomes critical (bottom
   right). Directly at $\Nf = 6$ the Thirring fixed point merges with
   fixed point $\mathcal I$ (bottom middle). In general, $\mathcal D$,
   $\mathcal H$, and $\mathcal I$ do not lie in the present plane---however, 
   we can observe their ``projections'' onto this plane at
   the points where the flow in directions parallel to this plane
   vanishes. These projections are marked by $\mathcal D'$, $\mathcal
   H'$, and $\mathcal I'$.}
 \label{fig:rg-flow-Thirring}
\end{figure*}
The large-$\Nf$ analysis~\cite{Parisi1975} shows the nonperturbative
renormalizability of the three-dimensional Thirring model with a
single interaction parameter $g_V$, which is equivalent to saying that
at large $\Nf$ the Thirring fixed point is critical, at least within
the $\mathrm U(2\Nf)$ subspace. In the context of the Thirring model's
lattice version, at large $\Nf$ it has been shown that small
perturbations that break the $\mathrm U(2\Nf)$ symmetry are RG
irrelevant and the $\mathrm U(2\Nf)$ is IR attractive, at least in the
vicinity of the Thirring fixed
point~\cite{DelDebbioHandsMehegan1997}. Our RG analysis is consistent
with the large-$\Nf$ behavior, but shows that the IR attractiveness of
the Thirring fixed point does \textit{not} reach all the way down to
$\Nf=2$.  Instead, there exists a ``critical'' number of flavors
$\Nf^{(2)}$ below which the Thirring subspace becomes IR repulsive and
the Thirring fixed point develops a second relevant direction. Below
$\Nf^{(2)}$, $\mathrm U(2\Nf)$-breaking perturbations are RG relevant
and drive the flow to a different fixed point with a lower
symmetry. 

This is visualized in Fig.~\ref{fig:rg-flow-Thirring}, which
shows the RG flow in the plane spanned by the two most relevant
directions at the Thirring fixed point $\mathcal C$, for different
number of flavors $\Nf$. Depending on where one starts the RG flow in
the vicinity of the Thirring fixed point one finds a finite region of
starting values for which the couplings flow to zero (Gaussian fixed
point). Outside this region there is a runaway flow. At the boundaries
of this region the flow runs into a critical fixed point with exactly
one relevant direction. Only for $\Nf > \Nf^{(2)} = 6$ is this the
Thirring fixed point itself. Below $\Nf^{(2)}$, these are either the
fixed points $\mathcal H$ and $\mathcal D$ for $2 \leq \Nf< \Nf^{(1)}$
or the fixed points $\mathcal I$ and $\mathcal D$ for $\Nf^{(1)} < \Nf
< \Nf^{(2)}$. Directly at $\Nf^{(2)}$ the Thirring fixed point merges
with $\mathcal I$.  We emphasize that $\Nf^{(2)}$ should \textit{not}
be confused with the chiral-critical flavor number
$N_\mathrm{f,c}^\chi$ below which the three-dimensional Thirring model
is expected to allow chiral symmetry
breaking~\cite{ChristofiHandsStrouthos2007, GiesJanssen2010,
  JanssenGies2012}. We believe that these two phenomena are unrelated
and the respective critical flavor numbers will most likely
\textit{not} coincide.

%%%%%%%%%%%%%%%%%%%%%%%%%%%%%%%%%%%%%%%%%%%%%%%%%%%%%%%%%%%%%%%%%%%%%%%%%%%%%%%%
\section{Prospects on long-range physics}
\label{sec:prospects-IR}
Within the present fermionic truncation of the effective average
action it is generically hard to associate a given critical fixed
point with a specific symmetry-breaking pattern and corresponding
continuous phase transition. Moreover, the general structure of the
beta functions as discussed in Sec.~\ref{sec:general-one-loop} renders
the single positive critical exponent $\Theta= {d-2}$, corresponding
to a correlation-length exponent of the associated phase transition of
$\nu = 1/\Theta = {1/(d-2)}$, independent of $\Nf$. While indeed many
fermionic universality classes in $d=3$, e.g., of the Gross-Neveu-type
\cite{RosensteinWarrPark1989, RosensteinYuKovner1993, Gracey1991,
  VasilievDerkachovKivelStepanenko1993, HandsKocicKogut1993,
  RosaVitaleWetterich2001, GiesJanssenRechenbergerScherer2010,
  BraunGiesScherer2011,
  JanssenHerbut2014,Borchardt:2015rxa,Vacca:2015nta} seem to point to
$\nu \approx 1 \ (\pm 20 \%)$, the insensitivity of the
fermionic-truncation prediction to the specific transition clearly
calls for more elaborate techniques to investigate the RG flow in the
vicinity of a given fixed point. Within the functional RG, this is,
for instance, possible by suitable partial or dynamical bosonization
techniques \cite{GiesWetterich2002, Pawlowski2007}. The nature of the
interacting phases expected at large coupling can also be investigated
by computing the flow of the order-parameter susceptibilities
\cite{SalmhoferHonerkampMetznerLauscher2004}, or the flows of
full potentials for fermion bilinears \cite{Aoki:2014ola}. Here, we
content ourselves with an outlook on \textit{possible} symmetry
breakings associated with the critical fixed points we have found and
leave a more detailed analysis for future work.

Some of the fixed points considered in the present work have been discussed 
earlier. Let us start with the Thirring fixed point $\mathcal C$. In the 
$\mathrm U(2\Nf)$ subspace, the Thirring fixed point is always a critical fixed 
point~\cite{GiesJanssen2010}. However, only for $\Nf = 1$ and $\Nf > \Nf^{(2)} = 
6 + \mathcal O(d-2)$ symmetry-breaking perturbations are RG irrelevant. For $\Nf 
= 1$ the three-dimensional Thirring model is expected to exhibit a spontaneous 
breaking of the ``chiral'' $\mathrm U(2\Nf)$ symmetry at large coupling $-g_V > 
-g_{V,c} \simeq - c_2(\Nf) > 0$ with order parameter 
$\langle\bar\psi\psi\rangle$. The critical behavior has been discussed within a 
functional RG approach~\cite{Mesterhazy2012}, yielding the following predictions 
for correlation-length critical exponent $\nu$ and order-parameter anomalous 
dimension $\eta$
\begin{align}
 \mathcal C(\Nf = 1): && \nu & \approx 1.9 & \eta & \approx 1.0.
\end{align}
For $\Nf>6$, where the Thirring fixed point again becomes critical
also in the presence of $\mathrm U(2\Nf)$-breaking interactions, the
ordering presumably no longer breaks the chiral
symmetry~\cite{JanssenGies2012}, but the exact type of ordering is
unknown.  The ordering is also not clearly identifiable for the
Thirring fixed point's equivalent for $g_V \leftrightarrow g_T$ for
$\Nf > 6$, i.e., the fixed point $\mathcal K$.

By contrast, the fixed point and the associated critical behavior of the 
three-dimensional Gross-Neveu model in the irreducible representation of the 
Clifford algebra ($\mathcal A$ in our notation) is fairly well known 
\cite{RosensteinWarrPark1989, RosensteinYuKovner1993, Gracey1991, 
VasilievDerkachovKivelStepanenko1993, HandsKocicKogut1993, 
RosaVitaleWetterich2001}. As expected from the location of the fixed point on 
the $(\bar\psi\gamma_{35}\psi)^2 = (\bar\chi\chi)^2$ axis, a nonvanishing order 
parameter $\langle \bar\psi\gamma_{35}\psi \rangle  = \langle \bar\chi \chi 
\rangle \neq 0$ occurs for large coupling $g_P > g_{P}^*|_\mathcal A \equiv a$ 
and thus breaks the parity symmetry while leaving the $\mathrm U(2\Nf)$ symmetry 
intact. Here again, $\psi$ corresponds to $\Nf$ flavors of reducible 
four-component Dirac spinors and $\chi$ to the corresponding $2\Nf$ 
two-component (Weyl) spinors [see Eq.~\eqref{eq:P2-barchi-chi}]. In the system 
of interacting fermions on graphene's honeycomb lattice the phase transition 
into the quantum anomalous Hall state that is predicted for large 
next-to-nearest neighbor interactions (at least for $\Nf = 1$ 
\cite{RaghuQiHonerkampZhang2008}) is expected to be governed by fixed point 
$\mathcal A$~\cite{HerbutJuricicRoy2009}. The critical behavior (for the example 
of $\Nf = 2$) is determined by the exponents
\begin{align} \label{eq:nu-eta-Gross-Neveu}
 \mathcal A(\Nf = 2): && \nu & \approx 0.95 \dots 1.04, & \eta & \approx 0.70 \dots 0.78,
\end{align}
where the ranges indicate the different predictions obtained by $(4-\epsilon)$ expansion \cite{RosensteinYuKovner1993}, large-$\Nf$ expansion \cite{VasilievDerkachovKivelStepanenko1993},  Monte-Carlo simulations \cite{HandsKocicKogut1993}, and functional RG \cite{RosaVitaleWetterich2001}. A recent overview of the literature results can be found in Ref.~\cite{JanssenHerbut2014}.

The fixed point $\mathcal D$ of the Gross-Neveu model in the reducible
representation has been considered in
Refs.~\cite{HerbutJuricicRoy2009,
  BraunGiesScherer2011,Borchardt:2015rxa,Vacca:2015nta}. $\mathcal D$
determines the critical behavior of the discrete-chiral-symmetry
breaking with order parameter $\langle \bar\psi\psi \rangle$ which
becomes finite for $g_S > a$. On the honeycomb lattice, it has been
ascribed to the transition into the charge density wave phase that is
expected for large nearest-neighbor interaction~\cite{Herbut2006,
  HerbutJuricicRoy2009}. The critical behavior for $\Nf \geq 2$
coincides with the one in the irreducible Gross-Neveu
model~[Eq.~\eqref{eq:nu-eta-Gross-Neveu}], at least to the order that
the exponents have been computed so far~\cite{RosaVitaleWetterich2001,
  BraunGiesScherer2011, Borchardt:2015rxa, Vacca:2015nta}.\footnote{In
  fact, the RG flows of the reducible Gross-Neveu model studied in
  \cite{BraunGiesScherer2011,Borchardt:2015rxa,Vacca:2015nta} are in
  principle identical to those of the irreducible Gross-Neveu model
  \cite{RosaVitaleWetterich2001} within the truncations focusing on
  the dynamics of the bosonic order parameter considered so far. In
  view of our results for the larger theory space, the literature
  results for the reducible case for $\Nf=1$ should rather be
  considered as applying more appropriately to the irreducible case.}
However, we would like to emphasize again that $\mathcal D$ develops
\textit{two} relevant directions in the one-flavor case. There is
therefore a fundamental difference between the irreducible and
reducible Gross-Neveu models for $\Nf=1$: As long as only one
microscopic parameter of a model, e.g., for spinless fermions on the
honeycomb lattice, is tuned, only $\mathcal A$ describes a second
order phase transition, and it can be driven by increasing $g_P$ for
small $g_S$ and $g_V$. By increasing the microscopic coupling $g_S$
and keeping $g_P$ and $g_V$ small and, say, positive we might find a
phase transition, which, however, is \textit{not} governed by fixed
point $\mathcal D$ but instead by the critical fixed point $\mathcal
E$.

Within the present approximation it is hard to decide which order
parameter is induced at fixed point $\mathcal E$. A simplistic
approach often applied
\cite{MetznerSalmhoferHonerkampMedenSchoenhammer2012} is given by
keeping track of the ``amount of divergence'' of the various
condensation channels, in order to determine in which channel the
couplings diverges ``first''. In our one-loop flow, the line
connecting $\mathcal E$ and the Gaussian fixed point $\mathcal O$ is a
RG attractive one-dimensional subspace for $\Nf=1$ and positive
couplings. E.g., if we start the flow in the UV near the
reducible-Gross-Neveu axis at
$\boldsymbol{g} = (g_S, g_V, g_P)$ with $0 < g_V, g_P \ll g_S$ and $g_S$ 
above but close to the critical $g_S^* = 1/2$,
the couplings will always run to the $\mathcal{OE}$ axis before they
eventually diverge at a finite RG scale $t_0$. From the values of the
fixed-point couplings [Eq.~\eqref{eq:fixed-point-E}] we find the
following ordering of couplings for $t \to t_0$
\begin{align}
 \frac{g_S}{g_V} & \to 3, &
 \frac{g_P}{g_V} & \to 1.38, & g_S, g_P, g_V & \to \infty,
\end{align}
i.e., the coupling $g_S$ diverges ``fastest''. From this simplistic
analysis one might thus speculate that fixed point $\mathcal E$
governs a condensation in the $(S)^2$ channel with order parameter
$\langle \bar\psi \psi\rangle$.  That is, with increase of the
microscopic coupling $g_S$ the Gross-Neveu model with reducible,
four-component, Dirac spinors should exhibit a continuous phase
transition beyond which the discrete $\mathbbm Z_2$ chiral symmetry is
spontaneously broken, a prediction that is consistent with the
mean-field theory for a system with large $g_S (\bar\psi\psi)^2$
interaction and $g_S > 0$~\cite{BraunGiesScherer2011}. This, of
course, requires confirmation beyond the present analysis. In any
case, however, we believe there is no reason to expect the critical
fixed point $\mathcal E$ to exhibit the same critical behavior as
fixed point $\mathcal A$. In our approximation, the largest critical
exponent is $\Theta = (d-2) + \mathcal O((d-2)^2)$ at any critical
fixed point. Yet, already the second-largest exponents that can be
associated to the (universal) corrections to scaling do not coincide:
at fixed point $\mathcal A$ we obtain
\begin{align}
\mathcal A(\Nf = 1): && \omega & = 2(d-2) + \mathcal O\!\left((d-2)^2\right),
%\end{align}
\intertext{whereas for fixed point $\mathcal{E}$ we get }
%\begin{align}
\mathcal E(\Nf = 1): && \omega & = (3-\sqrt{5})(d-2) +
\mathcal O\!\left((d-2)^2\right).
\end{align}%
Beyond our approximation, one would expect already the leading
exponents, for instance $\nu$ or $\eta$, to receive different
corrections at the two inequivalent fixed points. If true, the
irreducible and reducible Gross-Neveu models as defined by fixed
points $\mathcal A$ and $\mathcal E$ thus represent an example of two
three-dimensional fermion systems which both show spontaneous breaking
of $\mathbbm{Z}_2$ symmetry, but differ in their corresponding
critical behavior. This touches a general issue on universality in
fermionic systems: What are the defining properties that determine a
specific universality class? Our results suggest that in fermionic
systems the symmetry of the order parameter and the dimension and
field content of the given system does not yet uniquely define the
critical behavior. Instead, additional ``spectator symmetries'' that
do not take part in the symmetry breaking pattern might also play a
decisive role. In our case, this is the $\mathrm U(2\Nf)$ versus
$\mathbbm Z_2 \times \mathrm U(\Nf) \times \mathrm U(\Nf)$ symmetry
that discriminates between the irreducible and reducible Gross-Neveu
models. We believe that these general questions on universality are an
interesting direction for future research, and the models presented
here constitute a suitable playground to study them.

The fixed point $\mathcal H$ that becomes critical for $2\leq \Nf <
\Nf^{(1)} \approx 3.8$ has the interesting property that for $\Nf =
3.5$ it is located exactly on the axis at which all four couplings
coincide, $g_S = g_P = g_V = g_T < 0$. By means of the Fierz
identities the system on this axis can be written with pure $(S^D)^2 =
(\bar\psi^a \psi^b)^2 + (\bar\psi^a \gamma_{35} \psi^b)^2$
interaction [cf.\ Eq.~\eqref{eq:fierz-sd-vd}], and positive coupling
$g_S^D = -2g_S > 0$. One would thus expect the $(S^D)^2$ condensation 
channel
to become critical at this fixed point, and this should remain true
also for other values of $\Nf$ not too far from $\Nf=3.5$. However,
the critical behavior of this system is unknown to us.

We are also not able to make any substantial comment on the IR
behaviors of the fixed points $\mathcal I$ and $\mathcal J$. Both are
critical fixed points for $\Nf^{(1)} < \Nf < \Nf^{(2)}$, but
cannot
be associated to one of the possible condensation channels in an
obvious way.

%%%%%%%%%%%%%%%%%%%%%%%%%%%%%%%%%%%%%%%%%%%%%%%%%%%%%%%%%%%%%%%%%%%%%%%%%%%%%%%%
\section{Conclusions}
\label{sec:conclusions}

Our study of a general class of relativistic fermion theories in
$2<d<4$ space-time dimensions with continuous chiral $\mathrm U(\Nf)
\times \mathrm U(\Nf)$ symmetry has revealed a network of RG fixed
points. In a unified framework, our description includes a number of
well-studied models such as various versions of the Gross-Neveu model
and the Thirring model. If persistent also beyond our simple pointlike
approximation, each fixed point defines its own universality class
thus facilitating both different continuum limits and corresponding
microscopic ``theories'' as well as a diversity of possible long-range
phenomena. 

We pay particular attention to those fixed points with only one RG
relevant direction. These are candidates for critical points of a second-order 
quantum phase transition. It is one of our main results that the
nature of these critical fixed points in the present model depends on
the number of flavors. While the fixed point $\mathcal{A}$
corresponding to the irreducible Gross-Neveu model is a critical fixed
point for any number of flavors, the fixed point $\mathcal{D}$ for the
reducible Gross-Neveu model is only a critical fixed point for
$\Nf\geq 2$ whereas it is not for $\Nf=1$. Most interestingly, the
Thirring fixed point $\mathcal{C}$ is a critical fixed point for
$\Nf=1$ and for $\Nf>6$. While all these results are in agreement with
the corresponding large-$\Nf$ analyses of these theories, they
indicate that both the microscopic as well as the long-range behavior
of these systems can change drastically as a function of $\Nf$. 

Within the limit of pointlike interactions, we have proven a number of
properties of invariant subspaces of the RG flow. While subspaces of
higher symmetry are always guaranteed to be invariant subspaces, the
structure of our RG flows allows to define further criteria that are
not necessarily related to higher symmetries. Subspaces with higher
symmetry can lead to the phenomenon of emergent symmetry in the
long-range physics if this subspace is not only invariant but also RG
attractive towards the IR. Again we found that the emergence of higher
symmetry can be a flavor-number dependent property. For instance, the
Thirring subspace of higher $\mathrm{U}(2\Nf)$ symmetry is attractive
only for $\Nf> \Nf^{(2)}$ and $\Nf=1$ in the vicinity of
the Thirring fixed point $\mathcal C$.

As our approach is equivalent to the one-loop
$(2+\epsilon)$-expansion, at the very least it establishes the
\emph{existence} of $\Nf^{(2)}$ and its accompanying qualitatively
different behavior for $\Nf < \Nf^{(2)}$.  One should expect that our
estimate for its value in $d=3$ within the present simple truncation
is subject to quantitative improvement beyond the pointlike
limit. Nevertheless, we find it interesting that our one-loop result
$\Nf^{(2)} = 6 + \mathcal O(d-2)$ appears to be near the number
$N^\text{lattice}_\text{f,c} = 6.6(1)$ at which the IR observables in
the lattice simulations show an abrupt change
\cite{ChristofiHandsStrouthos2007}. These simulations employ a lattice
formulation of fermions (staggered fermions) that microscopically
breaks parts of the $\mathrm U(2\Nf)$ symmetry. It is therefore of
decisive importance for the interpretation of the simulation results
whether these perturbations are relevant in the RG sense or not. Our
analysis suggests that the answer to this question might in fact
depend on the flavor number, with a large-$\Nf$ regime in which
perturbations are irrelevant and an intermediate-$\Nf$ regime in which
perturbations become relevant. The boundary is given by
$\Nf^{(2)}$. It should be interesting for future work to establish
whether indeed the abrupt change found at
$N^\text{lattice}_\text{f,c}$ in the simulations signals---instead of
an upper bound for chiral symmetry breaking---a change in the number of
relevant directions at the UV Thirring fixed point. Such an
interpretation would reconcile the seeming disagreement between
$N^\text{lattice}_\text{f,c}$ and the majority of the analytical
estimates for $N_\mathrm{f,c}^\chi$, which appear to be significantly
lower than
$N^\text{lattice}_\text{f,c}$~\cite{GomesMendesRibeiroSilva1991,
  JanssenGies2012}.  It could potentially also resolve the
contradiction between the critical behaviors found in the continuum
Thirring model and its lattice version, as discussed in
\cite{JanssenGies2012}.

For $\Nf = 1$, our theory space describes the system of interacting
spinless fermions on the honeycomb lattice, a simple model for
graphene. This model has thoroughly been investigated
previously~\cite{RaghuQiHonerkampZhang2008, HerbutJuricicRoy2009,
  wang2014, garciamartinez2013}. Within our RG approach, we rediscover
the simple mean-field phase diagram which exhibits besides the
semimetallic phase at weak coupling two gapped phases, which are
approached by tuning nearest- and next-nearest-neighbor interactions,
respectively, above a strong-coupling threshold.  The appropriate
order parameters are $\langle\bar\psi\psi\rangle$ (charge density wave
phase) and $\langle \bar\psi \gamma_{35} \psi\rangle$ (quantum
anomalous Hall phase), respectively, each indicating a spontaneous
breaking of a $\mathbbm Z_2$ symmetry.  We can associate these
second-order quantum phase transitions with their corresponding
critical fixed points.  The transition into the quantum anomalous Hall
phase is determined by the irreducible-Gross-Neveu fixed point
$\mathcal A$, located on the axis with pure
$(\bar\psi\gamma_{35}\psi)^2$, as one would naively also expect from
mean-field theory.  A similar expectation, however, could fail in the
case of the charge density wave transition. Our results indicate that
this transition is \emph{not} to be associated with the
reducible-Gross-Neveu fixed point $\mathcal D$, which is located on
the axis with pure $(\bar\psi\psi)^2$ interaction and for $\Nf = 1$
has two RG relevant directions. Instead, the charge density wave
transition may be governed by fixed point $\mathcal E$, at which all
short-range interactions become finite.

Our results also provide evidence for a new feature of universality:
whereas universality is conventionally considered as being governed by
the symmetry of the order parameter, the dimensionality and the number
of long-range interactions, our theory space appears to contain a
possible counter-example: for $\Nf=1$, there are two critical fixed
points in the theory space ($\mathcal A$ and $\mathcal E$ in our
notation) which are likely to be associated with a second-order phase
transition involving the breaking of a $\mathbbm Z_2$ symmetry. In the
irreducible Gross-Neveu model, the phase transition describes a
breaking of parity symmetry, whereas the reducible Gross-Neveu model
goes along with a breaking of a discrete chiral symmetry. In both
cases, we have a $\mathbbm Z_2$ order parameter (real scalar field)
and the same number of massless fermion modes near the phase
transition. The main conceptual difference of the two models lies in
the additional spectator symmetries, which remain intact across the
phase transition. More formally, the phase transition is related to
two inequivalent RG fixed points lying in different invariant
subspaces of the full theory space. Quantitatively, we observe that the
corresponding critical exponents of the fixed points differ. Whereas
our simple approximation yields differing exponents only for the
subleading exponents, there is a priori no reason why an improved
approximation should not lead to differences also for the leading
relevant exponent.  If this is the case for the phase diagram of
spinless fermions on the honeycomb lattice, this would implicate that
the two possible phase transitions lie in different universality
classes. Even though both transitions would go along with the breaking
of a discrete $\mathbbm Z_2$ symmetry, they should exhibit a different
set of critical exponents.  The recent advances in overcoming the sign
problem in lattice Monte Carlo simulations~\cite{huffman2014} may
allow to test this prediction in the near future.

\acknowledgments{We have substantially benefited from discussions with
  J.\ Borchardt and I.\ F.\ Herbut. This work has been supported by
  the DFG under Grants FOR\,723, GRK\,1523, and JA\,2306/1-1, as well
  as the NSERC of Canada.}

%===BIBLIOGRAPHY ===========================================

%==================================================

%%%%%%%%%%%%%%%%%%%%%%%%%%%%%%%%%%%%%%
\end{document}